\documentclass[onecolumn]{emulateapj}
\usepackage{graphicx}
\usepackage{times}
\usepackage{verbatim}
\usepackage{natbib}
\usepackage{amsmath} 
\usepackage{amsbsy}

\slugcomment{}

\shorttitle{}
\shortauthors{}

\begin{document}

\title{Quasi-Resonant Theory of Tidal Interactions}

\author{Elena D'Onghia\altaffilmark{1,*}, Mark Vogelsberger\altaffilmark{1},
Claude-Andre Faucher-Giguere\altaffilmark{1,2,$\dagger$}, Lars Hernquist\altaffilmark{1}
}

\altaffiltext{1}{Harvard-Smithsonian Center for Astrophysics,
60 Garden Street, Cambridge, MA 02138 USA}
\altaffiltext{2}{Department of Astronomy,
University of California, Berkeley, CA 94720-3411, USA}
\altaffiltext{*}{Keck Fellow; edonghia@cfa.harvard.edu}
\altaffiltext{$\dagger$}{Miller Fellow}

\begin{abstract}

\noindent

When a spinning system experiences a transient gravitational encounter with an
external perturber, a \emph{quasi-resonance} occurs if the spin
frequency of the victim matches the peak orbital frequency of the
perturber.  Such encounters are responsible for the formation of long
tails and bridges of stars during galaxy collisions.  For high-speed
encounters, the resulting velocity perturbations can be described
within the impulse approximation.  The traditional impulse
approximation, however, does not distinguish between prograde and
retrograde encounters, and therefore completely misses the resonant
response.  Here, using perturbation theory, we compute the
effects of quasi-resonant phenomena on stars orbiting within a disk.
Explicit expressions are derived for the velocity and energy change to
the stars induced by tidal forces from an external gravitational
perturber passing either on a straight line or parabolic orbit.
Comparisons with numerical restricted three-body calculations
illustrate the applicability of our analysis.
\end{abstract}


\section{Introduction}

Astrophysical objects dominated by disks supported against gravity by
rotation are ubiquitous.  They range from planetary rings, to
planetary systems, to protoplanetary disks, to accretion disks around
young stars and compact objects, to spiral galaxies.  Given that
astrophysical disks are common, it is not surprising that
gravitational encounters involving them and other objects occur
frequently.  In particular, the numerous examples of peculiar galaxies
\citep[e.g.,][]{Arp66, Arp87} suggest that tidal interactions between
spirals are a key driver of the morphological transformations of
galaxies.

Beginning in the 1960s, various studies established that many peculiar
galaxies are, in fact, disks in collision
\citep{Pfle61,Yabu71,CB72a,CB72b,Wright72,E73}.  However, it was
\cite{TT72} (hereafter, TT72) who performed the first systematic study
of this process, using a restricted three-body technique.  In their
approach, each galaxy was modeled as a point-particle, representing
the potential well, surrounded by a disk of non-interacting test
particles (stars) on circular orbits.  When two such model galaxies
pass by one another, their mutual gravitational tidal forcing distorts
the disks as traced by the orbital motion of the test particles in the
combined potential of the two point-particles.  In this manner, TT72
showed that, contrary to prevailing wisdom, the narrow bridges and
tails seen in many peculiar galaxies could be produced by gravity
alone, and argued that these features are essentially kinematic in
nature.

TT72's analysis further demonstrated that the bridges connecting pairs
of colliding galaxies and the tidal tails that develop continue to
lengthen and thin out after an encounter.    The well-known systems NGC 4038 (the
``Antennae''), NGC 4676 (the ``Mice''), and NGC 7252 (the ``Atoms for
Peace Galaxy'') all have tails extending 50-100 kpc in length
\citep{Hib95}.  The Superantennae (IRAS19254-7245) is an extreme case
in which the tails span 350 kpc from tip to tip \citep{MLM91}.
If the galaxies merge, some
tail stars may escape the system entirely, but most eventually fall
back into the merger remnant, resulting in the formation of
``shells'' and other fine structures associated with elliptical
galaxies \citep{Malin80} through ``phase-wrapping''
\citep{Quinn84,HQ87,HS92}.

TT72 also showed that the efficiency of ``tail-making'' depends
sensitively on the inclinations of the disks relative to the orbit
plane.  This is illustrated in Figure 1, which shows the outcome
of two of TT72's simulations in which disks are perturbed by the
passage of equal-mass companions on parabolic orbits.  
The collisions are co-planar, meaning that the victim disks lie
exactly in the orbit plane, but in the top row of  Figure 1 the encounter is prograde,
so that the internal spin of the disk is aligned with the direction of
the orbital angular momentum, whereas in the bottom row the two are
anti-parallel, defining a precisely retrograde interaction.

\begin{figure}
\epsscale{1.0}
\plotone{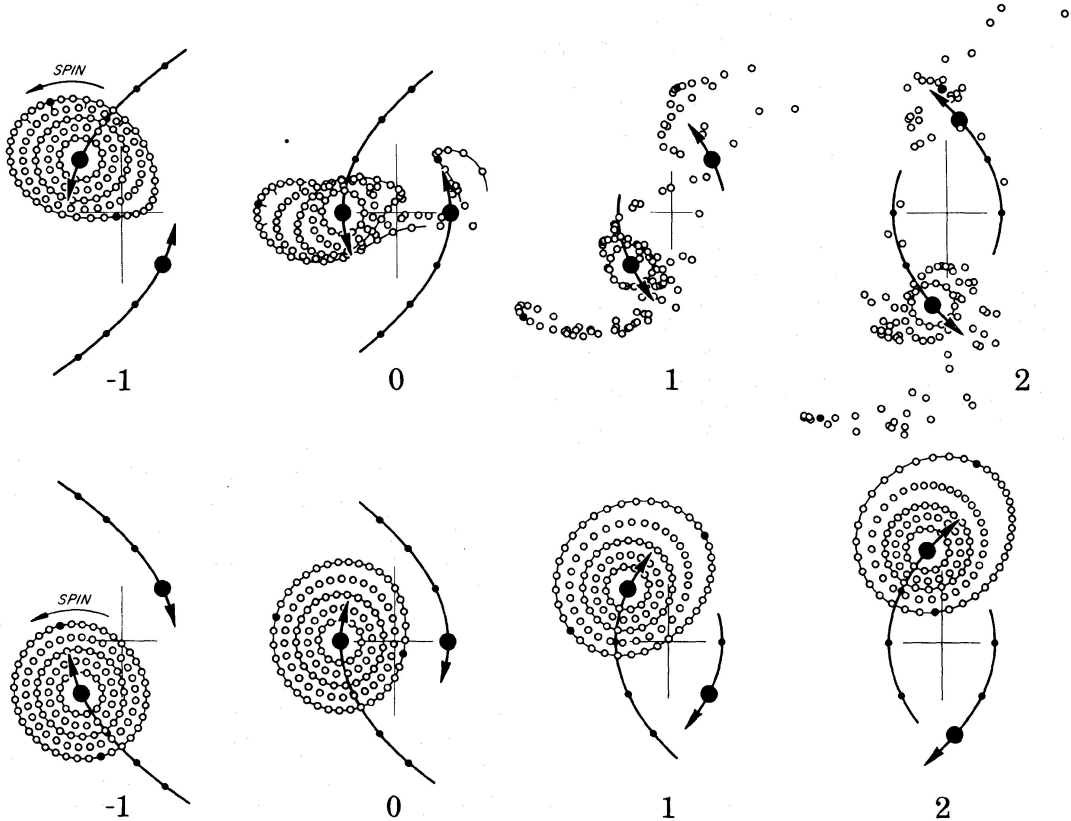}
\caption{Tail-making during a prograde encounter between two equal-mass
galaxies, represented by point-particles is shown in top row.  For clarity, the test particles
comprising only one of the disks are plotted.  In prograde collisions, the stars
in the disk are in near-resonance with the perturber during the interaction
and are continuously pulled either inward or outward, depending on their
internal orbital phases in relation to the orbital motion of the perturber.
A retrograde version of the encounter is shown in the bottom row. 
In this case, the stars in the disk are pulled alternately inward and outward during the
collision, with little net effect.
\citep [Adapted from][]{TT72} 
\label{Toomrepro}}
\end{figure}


It is obvious from visual inspection of Figure 1 that prograde
encounters can do much more violence to spinning disks than retrograde
ones.  TT72 interpreted this difference physically as owing to a
``near-resonance or matching of their [internal] orbital speeds with
the peak angular motion of the companion.''  In other words, a
relatively strong response follows if the spin angular frequency of
the disk is aligned with and of similar magnitude to the orbital
angular frequency of the collision at pericenter.  If the stars in the
disk are on circular orbits, but the trajectory of the interaction is
non-circular (as in the examples of parabolic orbits shown in Figures
1), then the orbital angular frequency varies with time and the
resonance is only temporary; hence, a ``near-'' or ``quasi-resonance.''
Mathematically, the condition for a strong response is expressed by
\begin{equation}
\Omega_{\rm{disk}} = \Omega_{\rm{orb}} \ \ \rightarrow \ \ 
\frac{v}{r} \sim \frac{V}{R \sqrt{(1+e)}},
\end{equation}
where $v$ and $r$ are the internal rotation velocity and
characteristic size of the disk, respectively, $V$ and $R$ are the
orbit velocity and separation at pericenter, and $e$ the eccentricity
of the orbit \citep{BT87}.

The work of \citet{B88,B92} and \citet{H92,H93}
generalized TT72's modeling by treating the internal gravity of
colliding galaxies self-consistently.  These and related studies
verified TT72's claim that tail-making is a kinematic process \citep[see
Figure 3 of][]{DMH99} and examined the
sensitivity of tail-making efficiency to the distribution of stars in
interacting and merging galaxies.

Subsequently, \citet{DMH96} and \citet{MDH98} demonstrated empirically that the
lengths and kinematics of tidal tails also depend critically on the
distribution of dark matter surrounding each galaxy.  For the same
radial stellar profiles, longer (shorter) and more (less) prominent
tidal tails form in shallower (deeper) dark matter potential wells.
In particular, \citet{DMH99} and \citet{SW99} showed that the depth of the dark
matter potential is as important in determining the lengths of tidal
tails as the orbit \citep[see e.g. Figure 4 in][]{DMH99}. Similar conclusions 
have been reached in studies of local dwarf spheroidal galaxies \citep [see e.g.]
[]{M02, Read06}.

Most recently, it has been shown that resonant stripping of stars in
disks can alter the mass to light ratios of dwarf galaxies when they
encounter more massive systems by removing luminous material more
efficiently than dark matter.  The minimal response of the dark matter
is expected if its particles move on random orbits, in which case the net
perturbation on the halo mostly averages out \citep{Don09b}.  A
similar resonant phenomenon has been suggested to cause the LMC disk
to thicken by interactions with the Milky Way \citep{W00} or more
generally in the context of heating and disruption of satellites
\citep{CWK09}. In addition, the origin of the Magellanic Stream might
have a tidal origin \citep{B10} through 
interactions between dwarfs in groups  \citep{DL08}, provided that 
the Magellanic Clouds are on their first pericentric passage \citep []{K06, B07}.

While these works have elucidated the physics of tail-making and
proven that observed peculiar galaxies are indeed a consequence of
collisions and mergers, they have left open a number of important
issues.  Most previous studies of tidal interactions between disk
galaxies have been empirical in nature.  Thus, they have identified
broad conditions for tidal features to be produced, but have not
provided simple criteria for isolating the efficiency of this 
process on specific
parameters of an encounter.  Therefore, the circumstances under which very
long tails could result, like those in the Superantennae, remain
uncertain.  Moreover, even nearly 40 years after TT72, there is still
considerable confusion in the community regarding galactic bridges and
tails, especially in regards to the role played by resonances in their
origin.  Finally, attempts to reproduce the morphology and kinematics
of individual observed systems is greatly complicated by the
sensitivity of tail-making to the distribution of dark matter
around galaxies.

One approach for dealing with this last issue is to survey the vast
parameter space that allows for variations in the orbit, the relative
masses and spatial sizes of the luminous galaxies, and the dark matter
potentials.  A promising scheme along these lines has been developed
recently by \citet{BH09} with their Identikit algorithm.  In what
follows, we pursue a complementary technique by formulating a simple
analytic description of tail-making to understand it from an alternate
perspective.  Our methodology necessarily entails various
approximations and is hence less general than simulation-based
procedures, but makes it possible to identify scalings of, and
interpret physically, the response of rotating objects to gravitational
tidal perturbations.  While our immediate attention is devoted to
galactic disks, our formalism is general with respect to the system
under consideration, so we anticipate that our analysis will be
relevant to a wide range of tidal phenomena.

In \S 2, we describe our method for analyzing tidal responses of
rotating systems.  We adopt a variant of the impulse approximation in
which we allow particles in the perturbed system to move along their
internal, unperturbed orbits during the encounter.  In \S 3, we
derive analytic expressions in terms of special functions for the
velocity perturbations delivered to spinning disks in these
interactions, in the case of coplanar collisions.  We do this first
for high-speed, straight-line trajectories, in \S 3.1, and then for
parabolic orbits in \S 3.2, the latter based partly on the work of
\citet{PT77} for tidal interactions between stars.  We present
numerical examples and contrast the straight-line and parabolic
cases in \S 3.3.  In \S 4, we generalize our formalism to
non-coplanar encounters, for both straight-line and parabolic orbits.
We compare the results of our analytic prescriptions to simulations
of tidal interactions in \S 5.  Finally, we summarize and conclude
in \S 6.

\section{Methodology}

Our approach is based on a variant of the impulse approximation for
studying gravitational perturbations on systems.  In some respects,
our formalism has similarities to that of \citet{PT77}, who considered
tidal interactions between stars as a means for forming close
binaries.  \citet{PT77} calculated the response of gas spheres to
external gravitational perturbations and the energy deposited into
non-radial oscillations.  Later, similar analytic estimates of the energy and
angular momentum exchange between a circumstellar disk and a passing
star on a near-parabolic orbit were inferred by \citet{O94} in the
context of accretion disks.  

Consider a flat, rotationally supported disk of stars, perturbed
gravitationally by a passing object.  In the usual impulse
approximation \citep{BT87}, it is assumed that the stars in the
perturbed system remain strictly stationary during the course of the
encounter.  The tidal force from the perturber relative to the center
of mass of the perturbed body is calculated at each location within it
from each point along the relative orbit of the interaction.  The
total velocity impulse delivered to each star in the perturbed object
is then calculated by integrating the force over the entire orbit.
In the simplest application of this method, the perturber follows a
straight-line trajectory, as for a high-speed encounter, although as
we show in what follows, it is possible to generalize the technique
also to parabolic collisions, following \citet{PT77}.

The analytic expressions that result from the impulse approximation
\citep{Spitz58} give reasonably accurate results for the energy
deposited in objects during tidal encounters for systems that are
dominated by internal random motions \citep{GO72,DSL80,AW85,Don10}.
However, this method is not appropriate for capturing the essentials
of responses like those in Figures 1 because, by construction,
it does not distinguish between prograde and retrograde interactions,
since the stars in the perturbed body are held fixed during the
encounter.

To qualitatively capture the influence of resonances during tidal
interactions, we employ the following variant of the impulse
approximation.  During the course of the encounter, stars in the
perturbed system are allowed to move along unperturbed orbits within
their host.  For simplicity, we assume that the unperturbed orbits are
strictly circular, although modest departures from these paths could,
in principle, be handled using epicyclic theory.  In this manner, the
response of a given star will depend not only on its spatial location
within the perturbed system, but also its velocity.  As we demonstrate
explicitly below, this approach makes it possible to characterize the
resonant aspects of tidal interactions that distinguish between
prograde and retrograde collisions, as in Figure 1.

If the orbit of the encounter is non-circular, as in the examples that
follow, the orbital angular frequency varies with time and, so, a
given star within the perturbed system will be in precise resonance
with the motion of the perturber for only a limited time.  For this
reason, we will refer to the formalism herein as describing 
{\it quasi-resonant} behavior, meaning that the response only resembles
that characteristic of a true resonant interaction.  This meaning
should be taken to be equivalent to TT72's description of tidal
interactions between spinning objects as displaying ``near-resonant''
qualities.

To be specific, consider a disk of stars on circular orbits
comprising the {\it victim} interacting with an object which we
will refer to as the {\it perturber}.  Employ a coordinate
system with origin at the center of mass of the victim and assume
that the disk is razor thin and orient the coordinate 
system so that the disk is in the $x-y$ plane.  Then,
the coordinate vector to any star within the victim is given by
\begin{equation}\label{vicorb}
\overrightarrow{r}(t)= (x(t),y(t),0) = (r \cos \phi(t), r \sin \phi(t), 0) ,
\end{equation}
\noindent
where $r$ is constant, because we assume that the
unperturbed orbits internal to the victim are circular,
and the position vector to the perturber will be denoted by
\begin{equation}
\overrightarrow{R}(t) = (X(t), Y(t), Z(t)) \, .  
\end{equation}
\noindent
In the sections below, we will consider various choices for the
trajectory $\overrightarrow{R}(t)$, which will fix the time-dependence
of the components $(X(t), Y(t), Z(t))$.  We will adopt the
convention that the unperturbed disk always lies in the
$x-y$ plane.  Thus, for non-coplanar encounters, we will
incline the orbit plane defined by the trajectory
$\overrightarrow{R}(t)$ so that $Z(t)$ will be non-zero.

The acceleration of each star relative to that on the center
of mass of the perturbed body is
\begin{equation}
{{d \overrightarrow{v}}\over{dt}} = - \left [ \nabla \Psi \, - \,
{1\over M} \int \rho (\overrightarrow{r}^\prime ) \nabla
\Psi (\overrightarrow{r}^\prime ) d^3\overrightarrow{r}^\prime
\right ] \, ,
\end{equation}
\noindent
where $\Psi$ is the interaction potential between the 
perturber and the victim,
$M$ is the mass of the victim, and the integral is over the
density profile of this object.  Expand $\nabla \Psi$ about
the origin in a Taylor series using
\begin{equation}
\Xi (\overrightarrow{r} + \overrightarrow{a}) \, = \,
\sum_{n=0}^\infty \, {1\over{n!}} \left ( \overrightarrow{a}
\cdot \nabla \right )^n \, \Xi (\overrightarrow{r}) \, .
\end{equation}
\noindent
After algebra, taking into account that the origin is located
at the center of mass of the victim, the $k$-th component of
the acceleration becomes
\begin{equation}
{{d v_k}\over{dt}} = - \Biggl \lbrace \sum_j r_j \left ( {{\partial ^2 \Psi}
\over{\partial r_j \partial r_k}} \right ) _{\overrightarrow{r}=0}
\, + \, {1\over 2} \sum_l \sum_j \left [ \left ( r_j r_l \, - \, H_{jl} \right ) 
\, \left (
{{\partial ^3 \Psi}
\over{\partial r_l \partial r_j \partial r_k}}
\right ) _{\overrightarrow{r}=0} \right ] \, + \, \ldots \, \Biggr \rbrace
\, ,
\end{equation}
\noindent
where $H_{jl} \equiv I_{jl} / M$ and $I_{jl}$ is the moment of
inertia tensor \citep{BT87}:
\begin{equation}
I_{jl} \, \equiv \, \int \rho r_j r_l d^3 \overrightarrow{r} \, .
\end{equation}
\noindent

For simplicity, treat the force on each star in the disk from the
perturber as that from a point mass.  Then, the interaction
potential is
\begin{equation}
\Psi = - {{GM_{\rm pert}}\over{\left | \overrightarrow{r}(t) -
\overrightarrow{R}(t) \right |}}.
\end{equation}
\noindent
Performing the derivatives required in the above expression,
we obtain the acceleration of a particular star at a given
time.  The velocity impulse delivered by the encounter can
then be obtained by integrating over time.  Thus, the
leading term in the series is \citep{BT87}
\begin{equation}\label{impulse}
\Delta \overrightarrow{v} \, = \, - G M_{\rm pert} \,
\int_{-\infty}^{\infty} \left [ {\overrightarrow{r}\over{R^3}} \, - \,
3 {{\overrightarrow{R} (\overrightarrow{r} \cdot \overrightarrow{R})}
\over{R^5}} \right ] dt \, .
\end{equation}
Likewise, the next order correction term can be
written
\begin{equation}
\Delta \overrightarrow{v}_{corr} \, = \, - {3\over 2} G M_{\rm pert} \,
\int_{-\infty}^{\infty} {1\over{R^5}} \Biggl \lbrace
2 \overrightarrow{r} (\overrightarrow{r} \cdot \overrightarrow{R})
\, + \, \left ( r^2 - Tr({\bf H}) - 2 {\bf H} \right ) 
\overrightarrow{R} \, + \,
5 {{\overrightarrow{R}}\over{R^2}} \left [
(\overrightarrow{R} \cdot {\bf H} \overrightarrow{R}) -
(\overrightarrow{r} \cdot \overrightarrow{R})^2 \right ]
\Biggr \rbrace dt \, .
\label{correction}
\end{equation}
In principle, this procedure can be extended to even higher
order terms in the series.
Note that the terms in this last
equation are all $\sim O (r^2)$ since
the elements in the matrix ${\bf H}$ involve integrals over
squares of the internal coordinates of the victim, while
the terms in eq. (\ref{impulse}) are $\sim O (r)$.

In what follows, we will employ eq. (\ref{impulse}) as the starting
point for our analysis.  Unlike as in the usual impulse approximation
we will allow both $\overrightarrow{R}(t)$ {\it and}
$\overrightarrow{r}(t)$ to vary in time.  We will, however,
assume that the trajectory of the interaction, specified by
$\overrightarrow{R}(t)$, is prescribed (i.e. orbital decay is not
accounted for), and that the orbital motion within the victim,
set by $\overrightarrow{r}(t)$, is such that the stars follow
their unperturbed motions throughout the course of the
interaction.

\section{Coplanar Encounters}

To illustrate our approach, we begin by considering a coplanar
interaction between a perfectly thin, rotating disk of stars and a 
passing perturber.  As noted earlier, the origin of the 
coordinate system will be at the center of mass of the
victim and the coordinate system will be oriented so
that the disk lies in the $x-y$ plane.  While not general,
this case suffices to characterize the dynamics of such encounters; we
will generalize to non-coplanar collisions later.
Denote the mass of the victim disk by $M$ and that of the
perturber by $M_{\rm{pert}}$.

\subsection{Straight-line trajectory}

\begin{figure}[ht]
\begin{center}
\epsscale{0.55}
\plotone{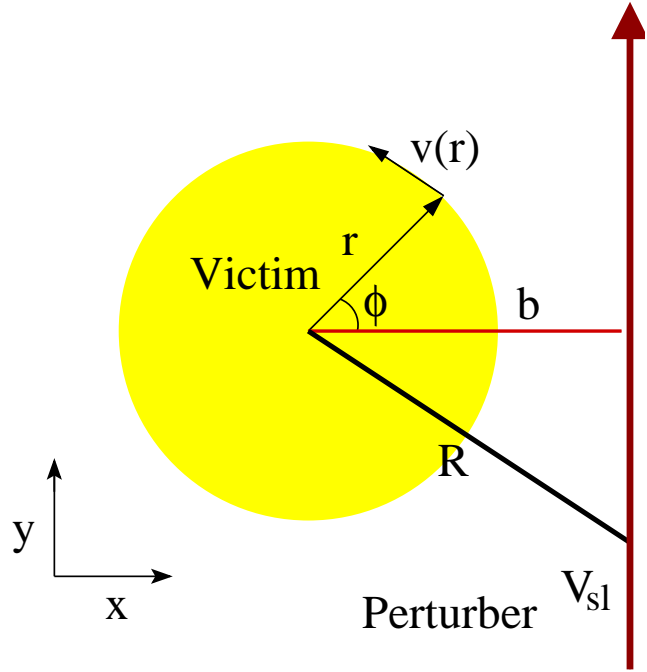}
\caption{Schematic view of a spinning disk galaxy (the victim) encountering a perturber passing 
by on a straight line at a constant speed $V_{sl}$. 
The stars in the victim are assumed to move on circular orbits.   
\label{Stripstraight}}
\end{center}
\end{figure}

The case of a perturber moving along a straight line relative to the
victim is the simplest one to analyze and is appropriate for
high-speed encounters, as in clusters of galaxies.  For definiteness,
take the orbit path to be
\begin{equation}\label{pathsl}
\overrightarrow{R}(t) = (b, V_{sl} \, t, 0) \, ,  
\end{equation}
\noindent
where, as indicated in Figure \ref{Stripstraight}, $b$ is the distance 
of closest approach (the impact parameter), which occurs at time
$t=0$, and $V_{sl}$ is the velocity of the encounter, which is constant
for a straight-line trajectory.  The internal motions of the disk
particles are given by eq. (\ref{vicorb}) with
\begin{equation}\label{phaset}
\phi(t)=\Omega t + \phi_0 \, ,
\end{equation}
where $\Omega$ the internal angular frequency of the victim and 
$\phi_0$ is the phase at the minimum distance from the perturber
at $t=0$.  Note that while $\Omega$ is assumed to be constant
in time, it can vary spatially according to
$\Omega = \Omega(r)$, depending on the shape of the 
rotation curve of the victim.
We will adopt the convention that $b$ and $V_{sl}$ are 
non-negative 
and distinguish prograde versus retrograde collisions 
for co-planar encounters by the sign
of $\Omega = \pm \mid \Omega \mid$ so that
\begin{eqnarray}
\Omega > 0 \ \ \ \ \ \ \ \ \  \rm{for} \ \rm{prograde} \ \rm{coplanar} \ \rm{encounters} \nonumber \ \ \ \ \ \ \ \ \ \ \ \ \ \ \ \\
 \ \Omega < 0 \ \ \ \ \ \ \ \ \  \rm{for} \ \rm{retrograde} \ \rm{coplanar} \ \rm{encounters} \, . \nonumber \ \ \ \ \ \ \ \ \ \ \ \ \\
\end{eqnarray}

\noindent
We further define the non-negative parameter $\alpha$ by
\begin{equation}\label{alphapar}
\alpha = \frac{\mid \Omega \mid b}{V_{sl}} \, .
\end{equation}
The orbital angular frequency varies with time along the trajectory
according to
\begin{equation}
\Omega_{orb} (R) \, = \, {{V_{sl}}\over R} = {{V_{sl}}\over {b (1+u^2)^{1/2}}} \, ,
\end{equation}
where $u=V_{sl} \, t / b$.  At the distance of closest approach, when $t=0$,
$\Omega_{orb} (b) = V_{sl} / b$ and so physically
\begin{equation}
\alpha = {{\mid \Omega \mid}\over{\Omega_{orb} (b)}}
\end{equation}
describes the condition for a resonance, with a strong 
response expected for $\alpha \sim 1$ when $\Omega > 0$.

Employing the usual trigonometric relations
\begin{equation}\label{trigcos}
\rm{cos}(\Omega t+\phi_0)= \rm{cos}\Omega t \cdot \rm{cos}\phi_0-\rm{sin}\Omega t \cdot \rm{sin}\phi_0
\end{equation}
\begin{equation}\label{trigsin}
\rm{sin}(\Omega t+\phi_0)= \rm{sin}\Omega t \cdot \rm{cos}\phi_0+\rm{cos}\Omega t \cdot \rm{sin}\phi_0
\end{equation}
and substituting into eq. (\ref{impulse}) gives

\begin{eqnarray}\label{coplxstru}
\Delta \rm{v}_x = - \rm{GM}_{\rm{pert}} \frac{\rm{r \ cos}\phi_0}{\rm{b}^2 \rm{V_{sl}}} \int_{-\infty}^{\infty}
\Big[\frac{\rm{cos}(\alpha \rm{u})}{(1+\rm{u}^2)^{3/2}}                                                                                        -3 \frac{\rm{cos}(\alpha \rm{u})}{(1+\rm{u}^2)^{5/2}} \mp
3 \rm{u}\frac{\rm{sin}(\alpha \rm{u})}{(1+\rm{u}^2)^{5/2}} \Big]\rm{du}
\end{eqnarray}

\begin{eqnarray}\label{coplystru}
\Delta \rm{v}_y= - \rm{GM}_{\rm{pert}} \frac{\rm{r} \sin \phi_0}{\rm{b}^2 \rm{V_{sl}}} \int_{-\infty}^{\infty}
\Big[-\frac{2\rm{cos}(\alpha \rm{u})}{(1+\rm{u}^2)^{3/2}}\pm   
3 \rm{u} \frac{\rm{sin}(\alpha \rm{u})}{(1+\rm{u}^2)^{5/2}} + 3 \frac{\cos (\alpha \rm{u})}{(1+\rm{u}^2)^{5/2}} \Big]\rm{du}
\end{eqnarray}

\noindent
and $\Delta v_z=0$.  In these expressions, for the terms with
$\mp$ or $\pm$ symbols the upper and lower signs correspond
to prograde and retrograde cases, respectively.

These integrals can be evaluated in terms of the modified 
Bessel functions $K_1$ and $K_2$ \citep{AS72}.
Employing the recursion relation
\begin{equation}
\alpha K_2(\alpha)=\alpha K_0(\alpha)+2 K_1(\alpha)
\label{besselrec}
\end{equation}
we arrive at
\begin{equation}
\Delta \rm{v}_x= 2 \rm{GM}_{\rm{pert}} \frac{\rm{r \ cos}\phi_0}{\rm{b}^2 \rm{V_{sl}}} 
\Big[\alpha K_1(\alpha) + \alpha^2\Big(K_0(\alpha) \pm K_1(\alpha)\Big)\Big] 
\label{coplxstrufinal}
\end{equation}

\begin{equation}
\Delta \rm{v}_y= -2 \rm{GM}_{\rm{pert}} \frac{\rm{r \ sin}\phi_0}{\rm{b}^2 \rm{V_{sl}}} \Big[\alpha^2 \Big(K_0(\alpha) \pm  K_1(\alpha)\Big)\Big], 
\label{coplystrufinal}
\end{equation}

\noindent
where the upper and lower signs are for 
prograde and retrograde encounters, respectively,
and $\Delta$v$_{z}$=0 for coplanar collisions.  The
structure of the expressions in terms of the modified Bessel functions
$K_0$ and $K_1$ is reminiscent of the results
describing the perturbations of orbits of stars within disks
owing to passing molecular clouds \citep{JT66}.

It is of interest to consider various limiting cases for these
expressions.  When $\alpha \rightarrow 0$, corresponding to
a slowly rotating system,
the Bessel
functions asymptote to
$K_0 (\alpha) \sim- \ln \alpha$ and $K_1(\alpha) \sim 1/ \alpha \ $
\citep{AS72}, and we find

\begin{equation}
\Delta \rm{v}_x \rightarrow 2 \frac{\rm{GM}_{\rm{pert}}}{\rm{b}^2\rm{V_{sl}}} \rm{r \ cos}\phi_0  \ \ \ \ \ \ \ \rm{for} \ \alpha  \rightarrow 0 
\end{equation}

\begin{equation}
\Delta \rm{v}_y \rightarrow 0 \ \ \ \ \ \ \ \ \ \ \ \ \ \ \ \ \ \ \ \ \ \ \ \ \ \ \ \ \ \ \ \ \ \rm{for} \ \alpha  \rightarrow 0
\end{equation}

\noindent
which can be obtained from the usual result for the impulse
approximation \citep[eq. (7-54) in][]{BT87}, as this limit
describes the situation when the stars in the disk remain
nearly stationary during the collision.  Note that this
limit is insensitive to the sign of $\Omega$ and hence 
does not distinguish between prograde and retrograde
encounters.

The limit $\alpha \rightarrow \infty$ corresponds to an  
interaction where the response should be weak because,
for example, the encounter is a distant one or the
spin and orbital frequencies are highly mismatched.
Employing the asymptotic expansion for the Bessel
functions \citep{AS72},
\begin{equation}
K_{\nu}(\alpha) \sim \frac{\sqrt{\pi}e^{-\alpha}}{\sqrt{2 \alpha}}
                     \Big[1+\frac{\mu -1}{8 \alpha}+\frac{(\mu -1)(\mu -9)}{2!(8\alpha)^2}+...\Big] \,,
\end{equation}
\noindent
where $\mu=4\nu^2$, we find for the prograde and retrograde cases
separately:

\begin{equation}
(\Delta \rm{v}_x)_{\rm{pro}} \sim 2 \frac{\rm{G}\rm{M}_{\rm{pert}}}{\rm{b}^2\rm{V_{sl}}} r \cos \phi_0 
\Big(\sqrt{2\pi} \alpha^{3/2} e^{-\alpha}\Big) 
\end{equation}

\begin{equation}
(\Delta \rm{v}_y)_{\rm{pro}} \sim -2 \frac{\rm{GM}_{\rm{pert}}}{\rm{b}^2\rm{V_{sl}}} r \sin \phi_0 
\Big(\sqrt{2\pi} \alpha^{3/2} e^{-\alpha}\Big) 
\end{equation}

\begin{equation}
(\Delta\rm{v}_x)_{\rm{retro}} \sim 2\frac{\rm{GM}_{\rm{pert}}}{\rm{b}^2\rm{V_{sl}}} r \cos \phi_0 
\Big(\frac{\sqrt{\pi}}{2^{3/2}} \alpha^{1/2} e^{-\alpha}\Big) 
\end{equation}

\begin{equation}
(\Delta\rm{v}_y)_{\rm{retro}} \sim 2\frac{\rm{GM}_{\rm{pert}}}{\rm{b}^2\rm{V_{sl}}} r \sin \phi_0 
\Big(\frac{\sqrt{\pi}}{2^{3/2}} \alpha^{1/2} e^{-\alpha}\Big).
\end{equation}

\noindent
We note that the response is exponentially suppressed in the limit
$\alpha \rightarrow \infty$, which demonstrates explicitly that
the perturbed system is protected by adiabatic invariance.  It is 
also interesting that in this limit the prograde and retrograde
cases are simply related:
\begin{equation}
\frac{(\Delta \rm{v}_x)_{\rm{pro}}}{(\Delta\rm{v}_x)_{\rm{retro}}} \sim 4 \alpha
\end{equation}

\noindent
and for the y-component:
\begin{equation}
\frac{(\Delta \rm{v}_y)_{\rm{pro}}}{(\Delta\rm{v}_y)_{\rm{retro}}} \sim - 4 \alpha \, .
\end{equation}

\noindent
Thus, the prograde response diverges relative to the retrograde one, by a factor
of $\alpha$, and, for a given $\alpha$ the magnitude in the velocity
perturbation is exactly a factor of four larger.

From the above expressions, the change in the energy of the 
perturbed system can be determined from:
\begin{eqnarray}\label{inputenergy}
(\Delta \rm{E})_{sl}=\frac{1}{2}\int \rho(\overrightarrow{r}) 
\mid \Delta \rm{v} \mid^2 d^3\overrightarrow{r} 
=\frac{2 \rm{G}^2\rm{M}_{\rm{pert}}^2}{\rm{b}^4\rm{V_{sl}}^2} \int_{0}^{\infty}r^3 dr \Sigma(r)
\cdot \nonumber \hspace{4cm}\\
\int_{0}^{2\pi}\Big[\Big(\alpha^2 K_0(\alpha)+
\alpha (1\pm \alpha)K_1(\alpha)\Big)^2 \cos^2 \phi 
+\alpha^4\Big(K_0(\alpha)\pm K_1(\alpha)\Big)^2 \sin^2 \phi \Big] d\phi \, , \ \ \ \ \ \ \ \ \ \ \ \ \
\end{eqnarray}
\noindent
where it is assumed that the disk is axisymmetric and has
zero vertical thickness, and $\Sigma(r)$ denotes the surface mass density distribution.  Doing the angular integral gives
\begin{eqnarray}
(\Delta \rm{E})_{sl}=\frac{2 \rm{G}^2\rm{M}_{\rm{pert}}^2 \pi}{\rm{b}^4\rm{V_{sl}}^2} 
\int_{0}^{\infty}\rm{r}^3 \rm{dr}\Sigma(r)\cdot \nonumber \hspace{8cm} \ \ \ \ \ \ \ \ \ \ \ \ \ \\
\Big[2 \alpha^4 K_0^2(\alpha)+2\alpha^3\Big(1\pm 2\alpha\Big)K_0(\alpha)K_1(\alpha)
+\alpha^2\Big(2\alpha^2\pm2\alpha+1\Big)K_1^2(\alpha)\Big] \, . \ \ \ \ \ \ \ \ \ \ \ \ \ \ \ \ 
\label{Ener_sl}
\end{eqnarray}

In the limit $\alpha \rightarrow 0$, the energy change is
\begin{eqnarray}\label{inputenalpha0}
(\Delta \rm{E})_{sl} \sim 2\pi \frac{\rm{G}^2\rm{M}_{\rm{pert}}^2}{\rm{b}^4\rm{V_{sl}}^2} \int_{0}^{\infty}
\rm{dr}  r^3 \Sigma(r) \, ,
\end{eqnarray} 
which agrees with the corresponding expression in \citet{BT87} using the
impulse approximation, as expected, while in the limit
$\alpha \rightarrow \infty$ for the prograde case
\begin{equation}
(\Delta \rm{E})_{\rm{pro}} \sim \frac{8\pi^2 \rm{G}^2\rm{M}_{\rm{pert}}^2}{\rm{b}^4\rm{V_{sl}}^2}
\int_{0}^{\infty}
\rm{dr}  r^3 \Sigma(r) \alpha^3e^{-2\alpha} 
\end{equation}
\noindent
and for a retrograde encounter
\begin{equation}
(\Delta \rm{E})_{\rm{retro}}\sim \frac{\pi^2 \rm{G}^2\rm{M}_{\rm{pert}}^2}{2\rm{b}^4\rm{V_{sl}}^2}
\int_{0}^{\infty}
\rm{dr}  r^3 \Sigma(r)  \alpha e^{-2\alpha} \, .
\end{equation}
To evaluate these, or eq. (\ref{Ener_sl}), it is necessary to specify the
radial dependence of $\alpha(r)$, i.e., the shape of the rotation
curve of the victim.

We have also evaluated the next order correction term given by 
equation (\ref{correction}), yielding: 
\begin{equation}
\Delta v^{corr}_{x} \, = \, - \frac{G M_{\rm pert} r^2} {b^3 V_{sl}} \,
\Biggr \lbrace -1 + \frac{2 H}{r^2} + {\rm cos} 2\phi_0 \Bigg[ -2 \alpha^2 \Big(1\pm 4 \alpha \Big) K_0(2 \alpha)
               -2 \alpha \Big(1\pm 2 \alpha + 4\alpha^2\Big) K_1(2 \alpha)\Bigg] \Biggr \rbrace
\label{impulse_corr_sl_x}
\end{equation}
\begin{equation}
\Delta v^{corr}_{y} \, = \, - \frac{G M_{\rm pert} r^2} {b^3 V_{sl}} {\rm sin} 2\phi_0 \cdot 2\alpha \,
\Biggr \lbrace \pm 4\alpha^2 K_0(2 \alpha) + K_1 (2 \alpha) \Big(4 \alpha^2 \pm \alpha \Big) \Biggr \rbrace,
\label{impulse_corr_sl_y}
\end{equation}
where the
upper and lower signs terms refer to
prograde and retrograde encounters, respectively,
and $\Delta$v$_{z} = 0$ because the encounter is assumed to be coplanar.
We note that the $H$ term does not couple to the phase $\phi_0$ of an
individual star, because it describes the overall reaction of the victim
to the perturber; i.e. the gain of angular momentum of the entire disk. 
On top of this global contribution, each star also receives a 
resonant sensitive
and $\phi_0$ dependent contribution. A similar effect occurs in linear order.
There, the overall contribution leads to a motion of the center of mass of
the victim. But since our calculation is performed in the center of mass
reference frame, this $\phi_0$ independent contribution does not show up
explicitly in the linear order results. There all terms depend on 
the phase $\phi_0$. We further note that the argument of the trigonometric
functions involving $\phi_0$ is different in first and second order:
$\phi_0$ in linear order and $2\phi_0$ in second order. This implies,
that the coupling of both orders leads to an asymmetric perturbation in
$x$ and $y$, whereas the linear term alone always produces symmetric results
as can be seen from the velocity increments in the $x$ and $y$ directions.

\subsection{Parabolic trajectory}

Next, consider an encounter from a parabolic trajectory, which is the
case analyzed by \citet{PT77} in their study of tidal interactions
between stars.  This situation is relevant for collisions between
galaxies in the field or in loose groups, where the orbits are highly
elongated.  For this reason, TT72 focused on this geometry in
particular, as in the examples shown in Figure 1.

The relative velocity between the victim and perturber in this case is
their mutual escape velocity, treating the interaction as that
between two point masses, and is given by:
\begin{equation}
\rm{V_p}(\rm{R})=\Big[\frac{2\rm{G}(\rm{M}_{\rm{pert}}+\rm{M})}{\rm{R}}\Big]^{1/2} \, .
\label{Vparab}
\end{equation}
\noindent
At the minimum separation between the victim and the perturber the 
relative velocity attains its maximum value V$_{0}$:
\begin{equation}
\rm{V_p}(\rm{b})=\rm{V}_0=\Big[\frac{2\rm{G}(\rm{M}_{\rm{pert}}+\rm{M})}{\rm{b}}\Big]^{1/2} 
\end{equation}
\noindent
and the relative velocity can be written as
\begin{equation}
\rm{V_p(R)}=V_0 \Big(\frac{b}{R}\Big)^{1/2} \, .
\end{equation}

\begin{figure}[ht]
\epsscale{0.5}
\plotone{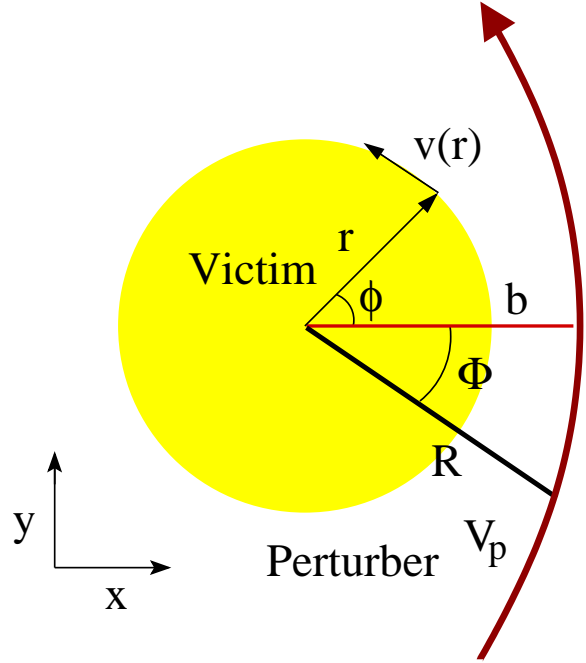}
\caption{Schematic view of a spinning disk galaxy (the victim) 
encountering a perturber on a parabolic orbit. 
\label{Stripparab}}
\end{figure}

We orient the disk as in the earlier derivation, so the orbits within
the victim disk are given again by eqs. (\ref{vicorb}) and
(\ref{phaset}).
To specify the orbit path, employ polar coordinates and write
\begin{equation}\label{polorb}
\overrightarrow{R}(t) = (X(t), Y(t), 0) =
(R(t) \cos \Phi (t), R(t) \sin \Phi (t), 0) \, .
\end{equation}
\noindent
The orbit is then specified parametrically by the
relations \citep{PT77}
\begin{equation}\label{moduloX}
\rm{R}=\rm{b}(1+\xi^2)
\end{equation}
and
\begin{equation}\label{xitan}
\xi=\rm{tan}\Big(\frac{\Phi}{2}\Big) ,
\end{equation}
so that the trajectory can be written as
\begin{equation}
Y^2=4b^2\Big(1-\frac{X}{b}\Big) \, .
\end{equation}
A schematic illustration of this case is shown in Figure \ref{Stripparab}.

The time-dependence of the orbit is given implicitly by
the relation
\begin{equation}\label{timeorb}
\rm{t}=\Big[\frac{2\rm{b}^3}
{\rm{G}(\rm{M}+\rm{M}_{\rm{pert}})}\Big]^{1/2}
(\xi+\frac{1}{3}\xi^3)=
\frac{2\rm{b}}{\rm{V}_0}(\xi+\frac{1}{3}\xi^3) \, .
\end{equation}

\noindent
We again introduce the parameter 

\begin{equation}
\alpha=\frac{\mid \Omega \mid b}{V_0},
\end{equation}
 
\noindent
with
$\Omega=\pm \mid \Omega \mid$, where the plus and minus signs are for 
prograde and retrograde coplanar encounters, respectively.
\noindent
The orbital angular frequency is
\begin{equation}
\Omega_{orb} (R) \, = \, {{V_p}\over R} = {{V_{0}}\over {b}} \left ( {b\over R} \right )^{3/2}  \, ,
\end{equation}
so that, as in the case of the straight-line trajectory
\begin{equation}
\alpha = {{\mid \Omega \mid}\over{\Omega_{orb} (b)}} \, ,
\end{equation}
and the largest response is expected for $\alpha \sim 1$.

To evaluate the velocity perturbations in this case, we begin with
eq. (\ref{impulse}) and adopt the time dependence of the motion within
the victim from eq. (\ref{vicorb}).  The phase angle of the stellar
orbits within the victim is given by
\begin{equation}\label{newphit}
\phi(t)=\pm 2 \alpha (\xi+\frac{1}{3}\xi^3)+\phi_0 \, .
\end{equation}
Expand $\cos \phi(t)$ and $\sin \phi(t)$ using trigonometric relations
analogous to eqs. (\ref{trigcos}) and (\ref{trigsin}) and 
parameterize the orbit of the encounter as in the polar form
of eq. (\ref{polorb}).  Finally, in the resulting integrals, make
a change of time coordinate from $t \rightarrow \xi$ according to
eq. (\ref{timeorb}).  Eliminating the integrals that are odd
in $\xi$ we find, after algebra:
\begin{equation}\label{deltavx1}
\Delta v_x = \frac{-4GM_{\rm{pert}} r \cos \phi_0 }{\rm{b}^2 \rm{V}_0} \int_{0}^{\infty}
\frac{\rm{d}\xi}{(1+\xi^2)^2}
\Biggl \lbrace \left (1 - 3 \cos^2 \Phi \right )
\cos\left [ 2\alpha(\xi+\frac{\xi^3}{3})\right ]
\, \mp \,
3 \sin \Phi \cos \Phi \sin \left [ 2\alpha(\xi+\frac{\xi^3}{3})\right ] \Biggr \rbrace
\end{equation}
\begin{equation}\label{deltavy1}
\Delta v_y = \frac{-4GM_{\rm{pert}} r \sin \phi_0 }{\rm{b}^2 \rm{V}_0} \int_{0}^{\infty}
\frac{\rm{d}\xi}{(1+\xi^2)^2}
\Biggl \lbrace \left (1 - 3 \sin^2 \Phi \right )
\cos\left [ 2\alpha(\xi+\frac{\xi^3}{3})\right ]
\, \pm \,
3 \sin \Phi \cos \Phi \sin \left [ 2\alpha(\xi+\frac{\xi^3}{3})\right ] \Biggr \rbrace .
\end{equation}
In these expressions, $\Phi$ depends implicitly on time
through eq. (\ref{timeorb}) according to
\begin{equation}
\Phi (\xi) = 2 \tan ^{-1} \xi
\end{equation}
and the upper and lower signs in the $\mp$ and $\pm$ terms refer to
prograde and retrograde coplanar encounters, respectively.

These can be put into a form amenable to further analysis by
repeated application of trigonometric relations of the form
\begin{equation}
\cos A \cos B \, = \, {1\over 2} \left [ \cos (A+B) + \cos (A-B) \right ] \, ,
\end{equation}
giving
\begin{equation}\label{deltavxp}
\Delta v_x = \frac{2GM_{\rm{pert}} r \cos \phi_0 }{\rm{b}^2 \rm{V}_0} \int_{0}^{\infty}
\frac{\rm{d}\xi}{(1+\xi^2)^2}
\Biggl \lbrace 
\cos\left [ 2\alpha(\xi+\frac{\xi^3}{3})\right ]
\, + \,
3 \cos \left [ 2\alpha(\xi+\frac{\xi^3}{3}) \mp 2 \Phi \right ] \Biggr \rbrace
\end{equation}
\begin{equation}\label{deltavyp}
\Delta v_y = \frac{2GM_{\rm{pert}} r \sin \phi_0 }{\rm{b}^2 \rm{V}_0} \int_{0}^{\infty}
\frac{\rm{d}\xi}{(1+\xi^2)^2}
\Biggl \lbrace 
\cos\left [ 2\alpha(\xi+\frac{\xi^3}{3})\right ]
\, - \,
3 \cos \left [ 2\alpha(\xi+\frac{\xi^3}{3}) \mp 2 \Phi \right ] \Biggr \rbrace.
\end{equation}

Finally, these results can be written in terms of the ``generalized'' Airy
functions of \citet{PT77}, who defined
\begin{equation}\label{Ilmpt77}
I_{lm}(y)=\int_{0}^{\infty}(1+\xi^2)^{-l}\cos [\sqrt{2}y(\xi+\frac{\xi^3}{3})+2m \  \rm{tan}^{-1} \xi]\rm{d}\xi,
\end{equation}
giving
\noindent
\begin{equation}\label{dvxpt77}
\Delta \rm{v}_x=\frac{2\rm{rGM}_{\rm{pert}} \cos \phi_0}{\rm{b}^2 \rm{V}_0}
\Big[I_{20}(\sqrt{2}\alpha)+3 I_{2\mp2}(\sqrt{2}\alpha)\Big]
\end{equation}
\begin{equation}\label{dvypt77}
\Delta \rm{v}_y=\frac{2\rm{rGM}_{\rm{pert}} \sin \phi_0}{\rm{b}^2 \rm{V}_0}
\Big[I_{20}(\sqrt{2}\alpha)-3 I_{2\mp2}(\sqrt{2}\alpha)\Big] \, ,
\end{equation}
where, as usual, the
upper and lower signs in the $\mp$ factors refer to
prograde and retrograde coplanar encounters, respectively.

Computation of these expressions is facilitated using the recursion
relations
\begin{equation}\label{Rec1}
I_{lm \pm 1}(\sqrt{2}\alpha)=\Big(2\mp \frac{2m}{l}\Big)I_{l+1,m}(\sqrt{2}\alpha)
-I_{lm}(\sqrt{2}\alpha)\mp \frac{2\alpha}{l}I_{l-1,m}(\sqrt{2}\alpha)
\end{equation}
\begin{equation}\label{Rec2}
I_{l0}(\sqrt{2}\alpha)=\frac{2l-3}{2l-2}I_{l-1,0}(\sqrt{2}\alpha)+
\frac{2\alpha^2}{(2l-2)(l-3)}I_{l-4,0}(\sqrt{2}\alpha) \, ,
\end{equation}
which are proven in \citet{PT77}.
Equation (\ref{Rec1}) makes it possible to obtain all the $I_{lm}$'s from the $I_{l0}$'s while 
equation (\ref{Rec2}) allows the $I_{l0}$'s for $l \geq 4$ to be computed 
from $I_{00},I_{10},I_{20},I_{30}$.
\noindent
In particular:
\begin{eqnarray}
\label{Delta vx parab}
I_{2\mp2}(\sqrt{2}\alpha)=-\frac{4}{3}I_{30}(\sqrt{2}\alpha)+(1\pm \frac{8}{3}\alpha)I_{20}(\sqrt{2}\alpha)
     \mp 2\alpha I_{10}(\sqrt{2}\alpha)+\frac{8}{3}\alpha^2 I_{00}(\sqrt{2}\alpha)
\end{eqnarray}
and, so
\begin{eqnarray}\label{impxparab}
\label{Delta vy parab}
\Delta \rm{v}_x=-\frac{4\rm{GM}_{\rm{pert}} r \cos \phi_0}{\rm{b}^2 \rm{V}_0} 
\Big[2 I_{30}(\sqrt{2}\alpha)-2(1\pm2\alpha)I_{20}(\sqrt{2}\alpha) 
\pm3\alpha I_{10}
 (\sqrt{2}\alpha)-4\alpha^2 I_{00}(\sqrt{2}\alpha)\Big] \ \ \
\end{eqnarray}

\begin{eqnarray}\label{impyparab}
\Delta \rm{v}_y=\frac{4\rm{GM}_{\rm{pert}} r \sin \phi_0}{\rm{b}^2 \rm{V}_0} 
\Big[2 I_{30}(\sqrt{2}\alpha)-(1\pm 4\alpha)I_{20}(\sqrt{2}\alpha) 
\pm 3\alpha I_{10}(\sqrt{2}\alpha)-4\alpha^2 I_{00}(\sqrt{2}\alpha)\Big], \  \ \ \ \ \ \ \ \ 
\end{eqnarray}

\noindent
where the
upper and lower signs in the $\pm$ terms refer to
prograde and retrograde co-planar encounters, respectively,
and $\Delta$v$_{z} = 0$.
For moderate arguments, the functions $I_{00},I_{10},I_{20},I_{30}$ can be computed by
numerical integration of eq. (\ref{Ilmpt77}) or from the
rational function approximations provided by \citet{PT77},
which are reproduced in the Appendix.

We again consider limiting cases of these
expressions.  When $\alpha \rightarrow 0$, corresponding to
a slowly rotating system,
it can be shown straightforwardly from the 
defining relation eq. (\ref{Ilmpt77}) that
$I_{20} \rightarrow \pi/4$ and
$I_{2\mp 2} \rightarrow 0$.  Thus, from
eqs. (\ref{dvxpt77}) and (\ref{dvypt77})
\begin{equation}
\Delta \rm{v}_x \rightarrow \frac{\pi}{2} \frac{\rm{GM}_{\rm{pert}} r \cos \phi_0}{\rm{b}^2 \rm{V}_0}
\end{equation}

\begin{equation}
\Delta \rm{v}_y \rightarrow \frac{\pi}{2} \frac{\rm{GM}_{\rm{pert}} r \sin \phi_0}{\rm{b}^2 \rm{V}_0} \, .
\end{equation}

\noindent
As earlier, this
limit is insensitive to the sign of $\Omega$ and hence 
does not distinguish between prograde and retrograde
encounters.

The change in energy of the victim in the
limit $\alpha \rightarrow 0$ is then
\begin{eqnarray}\label{enpar}
(\Delta \rm{E})_{par}=\frac{1}{2}\int \rho(\overrightarrow{r})\rm{d}^3\rm{r}\mid \Delta \rm{v} \mid^2 
=\frac{\pi^3 \rm{G}^2\rm{M}_{\rm{pert}}^2}{4 \rm{b}^4 V_0^2}
 \int_{0}^{\infty} \Sigma(r)\rm{r^3 \ dr} \, .
\end{eqnarray} 

\noindent
Comparing this change in energy to that produced by a perturber on a
straight-line trajectory, eq. (\ref{inputenalpha0}), we obtain
\begin{equation}\label{fracen}
\frac{\Delta \rm{E}_{\rm{par}}}{\Delta \rm{E}_{\rm{sl}}}= \frac{\pi^2}{8} \ \ \ \ \ (\rm{for} \, V_0=V_{\rm{sl}})\, .
\end{equation}

\noindent
Thus, if the velocity in the straight-line case is chosen to be
equal to the velocity at closest approach for a parabolic
encounter, the straight-line example provides a good
approximation to the damage done, at least when
$\alpha \rightarrow 0$.

The limit $\alpha \rightarrow \infty$ can be analyzed using
eqs. (\ref{impxparab}) and (\ref{impyparab}).  We note that in this
limit the rational function approximations provided by \citet{PT77}
for the functions $I_{00},I_{10},I_{20},I_{30}$ are not sufficiently
accurate to estimate the velocity perturbations for precisely
retrograde encounters because of exact cancellations between dominant
terms in the defining relations eqs. (\ref{impxparab}) and
(\ref{impyparab}).  Likewise, direct numerical integration of
eq. (\ref{Ilmpt77}) to sufficient accuracy for large values of
$\alpha$ is not possible because of the highly oscillatory behavior of
the integrand.  Instead, we employ asymptotic expansions for these
functions as $\alpha \rightarrow \infty$ obtained using the method of
steepest descent.  The results of this analysis are provided in
the Appendix. 
\cite{O94} previously obtained asymptotic expansions for the generalized Airy functions to leading order, but we carried out our analysis to higher orders since we consider limits in which the leading order terms cancel exactly.  

Combing eqs. (\ref{Delta vx parab})--(\ref{Delta vy parab}) with eqs. (\ref{I 0l infinity})--(\ref{I 3l infinity}), we find:

\begin{equation}
(\Delta \rm{v}_x)_{\rm{pro}} \sim 4 \frac{\rm{G}\rm{M}_{\rm{pert}}}{\rm{b}^2\rm{V_{0}}} r \cos \phi_0 
\Big(2 \sqrt{2\pi} \alpha^{3/2} e^{-4\alpha/3}\Big) 
\end{equation}

\begin{equation}
(\Delta \rm{v}_y)_{\rm{pro}} \sim -4 \frac{\rm{GM}_{\rm{pert}}}{\rm{b}^2\rm{V_{0}}} r \sin \phi_0 
\Big(2 \sqrt{2\pi} \alpha^{3/2} e^{-4\alpha/3}\Big) 
\end{equation}

\begin{equation}
(\Delta\rm{v}_x)_{\rm{retro}} \sim 4\frac{\rm{GM}_{\rm{pert}}}{\rm{b}^2\rm{V_{0}}} r \cos \phi_0 
\Big(\frac{1}{8}\sqrt{2\pi} \alpha^{1/2} e^{-4\alpha/3}\Big) 
\end{equation}

\begin{equation}
(\Delta\rm{v}_y)_{\rm{retro}} \sim 4\frac{\rm{GM}_{\rm{pert}}}{\rm{b}^2\rm{V_{0}}} r \sin \phi_0 
\Big(\frac{1}{8}\sqrt{2\pi} \alpha^{1/2} e^{-4\alpha/3}\Big) \, .
\end{equation}

\noindent
As for a straight-line trajectory, the response is exponentially suppressed
in the limit $\alpha \rightarrow \infty$, owing to adiabatic invariance.
The prograde and retrograde cases are again simply related:
\begin{equation}
\frac{(\Delta \rm{v}_x)_{\rm{pro}}}{(\Delta\rm{v}_x)_{\rm{retro}}} \sim 16 \alpha
\end{equation}

\noindent
and for the y-component:
\begin{equation}
\frac{(\Delta \rm{v}_y)_{\rm{pro}}}{(\Delta\rm{v}_y)_{\rm{retro}}} \sim - 16 \alpha \, .
\end{equation}

\noindent
This is similar to the result obtained earlier
for straight-line collisions, but now with a numerical
coefficient of 16 rather than four.

The corresponding changes in energy in the limit
$\alpha \rightarrow \infty$ are, for the prograde case
\begin{equation}
(\Delta \rm{E})_{\rm{pro}} \sim \frac{128\pi^2 \rm{G}^2\rm{M}_{\rm{pert}}^2}{\rm{b}^4\rm{V_{0}}^2}
\int_{0}^{\infty}
\rm{dr}  r^3 \Sigma(r) \alpha^3e^{-8\alpha/3} 
\end{equation}
\noindent
and for a retrograde encounter
\begin{equation}
(\Delta \rm{E})_{\rm{retro}}\sim \frac{\pi^2 \rm{G}^2\rm{M}_{\rm{pert}}^2}{2\rm{b}^4\rm{V_{0}}^2}
\int_{0}^{\infty}
\rm{dr}  r^3 \Sigma(r)  \alpha e^{-8\alpha/3} \, .
\end{equation}

\noindent
Quantitatively, the relation of these results to those for a straight-line
trajectory will depend in detail on the rotation curve of the victim 
through the radial dependence of $\alpha = \alpha (r)$.

We have also evaluated the next order correction term given by eq. (\ref{correction}),
yielding:
\begin{equation}
\Delta v^{corr}_{x} \, = \, - \frac{6 G M_{\rm pert} r^2} {b^3 V_{0}} \,
\Biggr \lbrace \frac{\pi}{8}\Big(-\frac{1}{2}+\frac{H}{r^2}\Big) \ +
      {\rm cos} 2\phi_0 \Big[-\frac{1}{4}I_{3\mp 1}(2\sqrt{2}\alpha) - \frac{5}{4} I_{3 \mp 3}(2\sqrt{2}\alpha) \Big] \Biggr \rbrace
\label{impulse_corr_par_x}
\end{equation}
\begin{equation}
\Delta v^{corr}_{y} \, = \, - \frac{6 G M_{\rm pert} r^2} {b^3 V_{0}} {\rm sin} 2\phi_0 \,
\Biggr \lbrace \Big[-\frac{1}{4}I_{3\mp 1}(2\sqrt{2}\alpha) + \frac{5}{4} I_{3 \mp 3}(2\sqrt{2}\alpha) \Big] \Biggr \rbrace, \  
\label{impulse_corr_par_y}
\end{equation}
where the
upper and lower signs in the $\pm$ terms refer to
prograde and retrograde co-planar encounters, respectively,
and $\Delta$v$_{z} = 0$.

\subsection{Straight line paths versus  parabolic passages}
We now compare the perturbations in velocities of stars in a thin disk
in the x-y plane owing to a coplanar tidal interaction with a system
passing on either a straight line path or a parabolic orbit.  To
isolate the dependence on the parameter $\alpha$, we define, for a
given relative velocity $V$, impact parameter $b$, and mass of the
perturber M$_{\rm{pert}}$ the quantities:

\begin{eqnarray}
(\Delta \rm{v}_x)'= \frac{\Delta \rm{v}_x}{\frac{\rm{GM}_{pert}}{\rm{b}^2\rm{V}} \rm{r}\cos \phi_0} 
\end{eqnarray}

\begin{eqnarray}
(\Delta \rm{v}_y)'= \frac{\Delta \rm{v}_y}{\frac{\rm{GM}_{pert}}{\rm{b}^2\rm{V}} \rm{r}\sin \phi_0}.                   
\end{eqnarray}

\noindent
For the case of encounters on straight line paths:
\begin{eqnarray}
(\Delta \rm{v}_x)_{\rm{sl}}^{'}=2[\alpha^2 K_0(\alpha)+\alpha(1\pm \alpha) K_1(\alpha)]
\end{eqnarray}

\begin{eqnarray}
(\Delta \rm{v}_y)_{\rm{sl}}^{'}=-2[\alpha^2 K_0(\alpha)\pm \alpha^2 K_1(\alpha)] \, ,
\end{eqnarray}

\noindent
while for parabolic encounters:
\begin{eqnarray}
(\Delta \rm{v}_x)_{\rm{par}}^{'}= -4[2I_{30}(\sqrt{2}\alpha)-2(1\pm2\alpha)I_{20}(\sqrt{2}\alpha)
                  \pm 3\alpha I_{10}(\sqrt{2}\alpha)-4\alpha^2 I_{00}(\sqrt{2}\alpha)]
\end{eqnarray}

\begin{eqnarray}
(\Delta \rm{v}_y)_{\rm{par}}^{'}= 4[2I_{30}(\sqrt{2}\alpha)-(1\pm 4\alpha)I_{20}(\sqrt{2}\alpha)
                  \pm 3\alpha I_{10}(\sqrt{2}\alpha)-4\alpha^2 I_{00}(\sqrt{2}\alpha)].
\end{eqnarray}

\begin{figure}[ht]
\epsscale{1.0}
\plotone{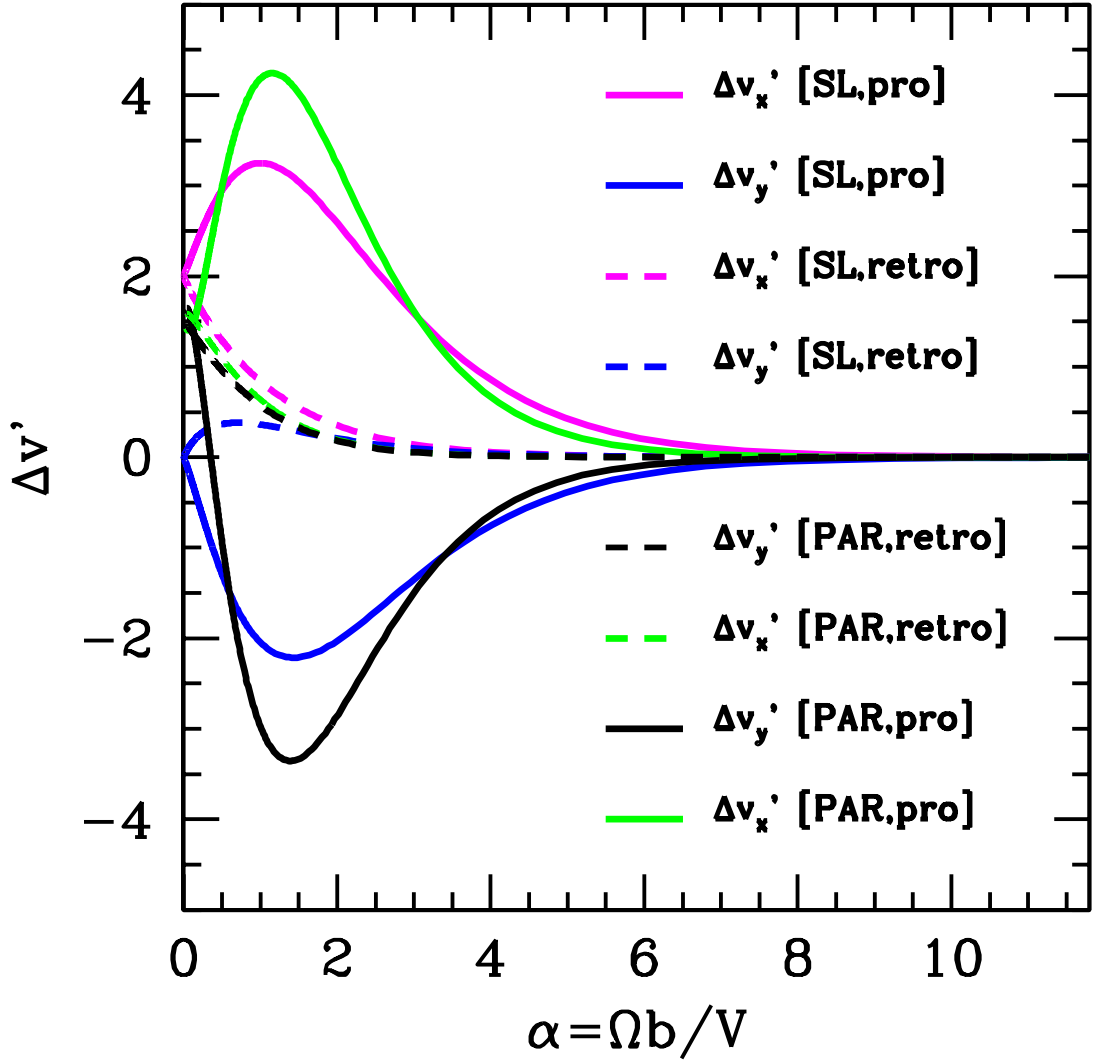}

\caption{Velocity increments of stars rotating in a victim 
while encountering a system on straight line or parabolic orbits.  Shown
are the quantities $(\Delta \rm{v}_x)_{\rm{sl}}^{'}$
and $(\Delta \rm{v}_y)_{\rm{sl}}^{'}$ as functions of the parameter
$\alpha$ for prograde and retrograde encounters.
\label{alpha}}
\end{figure}

Figure \ref{alpha} shows the velocity perturbations on stars within a spinning 
disk from either straight-line or parabolic encounters.
For the prograde cases, the response is a maximum near $\alpha \sim 1$, where there is a matching between the spin 
frequency of the rotating victim and the orbital frequency around the perturber,
\begin{equation}
\Omega_{\rm{victim}} = \Omega_{\rm{pert}} \ \ \rightarrow \ \ \frac{v}{r} \sim \frac{V}{R},
\end{equation}
\noindent
characteristic of a resonance.  However, the resonance is broad, with
FWHM $\Delta \alpha \sim 2$; hence, we refer to this phenomenon as 
a quasi-resonance.

The width of the resonance can be understood from the fact that, owing
to the non-circular nature of the perturber's orbit, its effective
orbital frequency as perceived by the victim changes with time.  As a
result, the resonance is ``maximal'' only for a short period of time.
Furthermore, at each point along the orbit the instantaneous orbital
frequency can be in resonance with stars located at different radii
within the victim, and can therefore act more efficiently on different
parts of it at different times.

In a true resonant interaction, it is assumed that a periodic
perturbation is applied for an indefinite length of time.  In that
event, the resonance will be narrow because perturbations
off-resonance will cancel out in the limit of infinitely many cycles.
For a tidal encounter from an unbound trajectory, as we consider here,
this is not necessarily the case and off-resonant driving forces can
result in a non-zero perturbation.  This incomplete cancellation
explains why the response for prograde encounters shown in Figure
\ref{alpha} is broad.  For retrograde interactions, the cancellation
is more complete and the response is suppressed.  For both prograde
and retrograde cases, in the limit $\alpha \rightarrow \infty$, a star
in the victim can complete many orbits while the perturber is near a
particular orbital frequency, resulting in an averaging out of the
perturbation, yielding an exponentially suppressed response.

Figure \ref{alpha} also illustrates the maximal response in the
different cases for $\alpha \sim$1.  Near this value of $\alpha$, the
velocity increments owing to the quasi-resonant phenomenon are greater
when the victim experiences a parabolic encounter than in the case of
a straight line path, since the curvature of the orbit implies that
the resonance is applied for a longer time.  For non-rotating systems
($\alpha \rightarrow 0$), the velocity increment gained by the stars
in the victim is described by the usual impulse approximation.  In
quasi-resonance, the velocity perturbations in the prograde cases are
a factor $\sim 2-3$ higher than in this limit, while the response is suppressed
of a similar factor for a retrograde collision for $\alpha \sim 1$.

Note that in our analysis, the disk is assumed to be infinitely thin. However 
the large resonance width shown in Figure \ref{alpha} would make the results applicable
even to rotating systems with a finite velocity dispersion.  Qualitatively we expect
that for a ``hot'' disk victim the resonant response would still occur,
although the efficiency will be somewhat reduced.

\section{Non-coplanar encounters}

In order to generalize the formulation presented earlier to
non-coplanar encounters, we need to consider situations where the
orbit plane is arbitrarily inclined relative to that of the victim
disk.  To facilitate interpretation of the results, especially
vertical heating of the disk, we choose to perform this operation by
actively rotating the orbit plane, leaving the unperturbed disk in the
$x-y$ plane.  Thus, as above, the coordinates of stars in the victim
disk are still given by eq. (\ref{vicorb}) and the trajectory of the
perturber, $\overrightarrow{R}(t)$, will be specified by rotating its
coplanar path, which we will refer to as
$\overrightarrow{R}_{cop}(t)$ in the following.

To describe the rotation, we introduce the 
$3 \times 3$~ matrix $\tilde{A}$,
given by:
\[ \tilde{A} = \left [ \begin{array}{ccc}
a_{11} & a_{12} & a_{13} \\
a_{21} & a_{22} & a_{23} \\
a_{31} & a_{32} & a_{33}  \end{array} \right ] . \] 
Then the rotated trajectory of the encounter is given
by
\begin{equation}
\overrightarrow{R}(t) = \tilde{A}
\overrightarrow{R}_{cop}(t) \, .
\end{equation}
Rotating the path of the perturber does not
alter the magnitude of $\overrightarrow{R}(t)$ and so
\begin{equation}
R(t) = \left | \tilde{A} \overrightarrow{R}_{cop}(t)
\right | = R_{cop}(t) \, .
\end{equation}

The convention for the Euler angles that specify the rotation
matrix $\tilde{A}$ is not unique.  

\noindent
Defining
\[ \begin{array}{cc}
c_{\lambda}=\cos \lambda;   s_{\lambda}=\sin \lambda \\
c_{\eta}=\cos \eta;         s_{\eta}=\sin \eta~, \\
c_{\chi}=\cos \chi;     s_{\chi}=\sin \chi \end{array}  \] 
we take the rotation matrix to be
\[ \tilde{A} = \left [ \begin{array}{ccc}
c_{\lambda}c_{\eta}c_{\chi}-s_{\lambda}s_{\chi}  &-c_{\lambda}c_{\eta}s_{\chi}-s_{\lambda}c_{\chi}  & c_{\lambda}s_{\eta}  \\
s_{\lambda}c_{\eta}c_{\chi}+c_{\lambda}s_{\chi}  &-s_{\lambda}c_{\eta}s_{\chi}+c_{\lambda}c_{\chi}  & s_{\lambda}s_{\eta} \\
-s_{\eta}c_{\chi}   & s_{\eta}s_{\chi} &  c_{\eta} \end{array} \right ], \]
corresponding to a first rotation about the $z$ axis by an angle $\chi$, a second rotation about the $y$ axis by an angle $\eta$, and a third rotation about the $z$ axis again by an angle $\lambda$. 
In other words, $\tilde{A} \equiv R(z,~\lambda) R(y,~\eta) R(z,~\chi)$, where $R(x_{i},~\theta)$ is a standard matrix describing a rotation by an angle $\theta$ about the $x_{i}$ axis. 
The angles $\chi$, $\eta$, and $\lambda$ can take arbitrary values (positive or negative), with the sense of rotation determined by the right hand rule. 
For example, the orbital angular momentum for a coplanar
collision is $\overrightarrow{L}_{cop}(t) = L_{orb} \hat{z}$,
implying
\begin{equation}
\overrightarrow{L}(t) = \tilde{A} \overrightarrow{L}_{cop}(t) =
L_{orb} (\cos \lambda \sin \eta , \sin \lambda \sin \eta , 
\cos \eta ) \, .
\end{equation}

\subsection{Encounters on straight line paths}
\noindent
We consider first the case of non-coplanar encounters in which
the perturber moves along a straight-line, generalizing the
results of \S 3.1.  The coplanar trajectory is given by
eq. (\ref{pathsl}), the phase angle of the stars along their
orbits in the victim is, as earlier, defined by
eq. (\ref{phaset}), and we employ the parameter $\alpha$,
as defined by eq. (\ref{alphapar}).  Note, however, that now
the sign of $\Omega$ does not determine if the encounter
is prograde or retrograde, but merely fixes the direction of
the spin angular momentum of the victim, either along the
$+z$ axis (positive $\Omega$) or 
$-z$ axis (negative $\Omega$).  For non-coplanar collisions,
the condition of whether
the encounter is mainly prograde versus retrograde is set
by $\beta = \overrightarrow{L}(t) \cdot \overrightarrow{S}(t)/
(L S)$, where 
$\overrightarrow{L}(t)$ and $\overrightarrow{S}(t)$ are the
orbital angular momentum of the encounter and the spin
angular momentum of the victim disk, respectively.  The
quantity $\beta$ lies between $+1$ (purely prograde) and
$-1$ (purely retrograde) and takes the value $\beta = 0$
for a polar orbit.

Starting from the expressions for the velocity
perturbations in eq. (\ref{impulse}) we find, after
algebra:

\begin{align}
\label{dvxnoncop}
\Delta v_x  =  - {{2GM_{pert}}\over{b^2 V_{sl}}} \, r \,
\Biggl \lbrace & \alpha ^2 K_0(\alpha) \left [
(a_{12}^2 - a_{11}^2) \cos \phi_0 +
(a_{12} a_{22} - a_{11} a_{21}) \sin \phi_0 \right ] \notag \\
\, & + \, \alpha K_1(\alpha) \left [
(1-2a_{11}^2-a_{12}^2) \cos \phi_0 -
(a_{12}a_{22} + 2 a_{11}a_{21}) \sin \phi_0 \right ]
\, \notag \\
\, & \mp \alpha^2 K_1(\alpha) \left [
(a_{11}a_{22} + a_{21} a_{12}) \cos \phi_0 -2a_{11}a_{12}
\sin \phi_0 \right ] \Biggr \rbrace
\end{align}

\begin{align}
\label{dvynoncop}
\Delta v_y = - {{2GM_{pert}}\over{b^2 V_{sl}}} \, r \,
\Biggl \lbrace & \alpha ^2 K_0(\alpha) \left [
(a_{12}a_{22} - a_{11}a_{21}) \cos \phi_0 +
(a_{22}^2 - a_{21}^2) \sin \phi_0 \right ]
\, \notag \\
 & + \, \alpha K_1(\alpha) \left [
(1-2a_{21}^2-a_{22}^2) \sin \phi_0 -
(a_{12}a_{22} + 2 a_{11}a_{21}) \cos \phi_0 \right ] \notag \\
\, & \pm \, \alpha^2 K_1(\alpha) \left [
(a_{21}a_{12} + a_{22} a_{11}) \sin \phi_0 -2a_{21}a_{22}
\cos \phi_0 \right ] \Biggr \rbrace
\end{align}

\begin{align}
\label{dvznoncop}
\Delta v_z  =  -{{2GM_{pert}}\over{b^2 V_{sl}}} \, r \,
\Biggl \lbrace & \alpha ^2 K_0(\alpha) \left [
(a_{32}a_{12} - a_{31}a_{11}) \cos \phi_0 +
(a_{32}a_{22} - a_{31}a_{21}) \sin \phi_0 \right ] \notag \\
\, & + \, \alpha K_1(\alpha) \left [ -
(a_{32}a_{22}+2a_{31}a_{21}) \sin \phi_0 -
(a_{32}a_{12} + 2 a_{11}a_{31}) \cos \phi_0 \right ] \notag \\
\, & \pm \, \alpha^2 K_1(\alpha) \left [
(a_{31}a_{12} + a_{32} a_{11}) \sin \phi_0 -
(a_{31}a_{22}+a_{21}a_{32})
\cos \phi_0 \right ] \Biggr \rbrace .
\end{align}

\noindent
Here, the factors $a_{ij}$ are the 
relevant components of the rotation matrix $\tilde{A}$,
and the upper signs in the $\mp$ and $\pm$ terms refer to
the case with $\Omega > 0$ while the lower signs are for the
case with $\Omega < 0$.
It is straightforward to show that these expressions
reduce to the appropriate ones for coplanar encounters
given in \S 3.1 if the rotation angles are set to zero.

As a simple illustration, consider a straight-line path
where the original orbit is in the $x-y$ plane and is
given by $\overrightarrow{R}_{cop}(t) = (b, V_{sl} \, t, 0)$,
and employ the Euler angle convention summarized above.
Rotate this path by an angle $\eta$ around the 
$y$-axis, so that
\begin{equation}\label{pathnoncopsl}
\overrightarrow{R}(t) = (b \cos \eta, V_{sl} \, t, - b \sin \eta) \, .  
\end{equation}
\noindent
In this case,
\begin{equation}
\overrightarrow{L}(t) =
L_{orb} ( \sin \eta , 0, \cos \eta ) \, ,
\end{equation}
and the components of the rotation matrix are
given by
\[ \tilde{A} = \left [ \begin{array}{ccc}
\cos \eta  & 0 & \sin \eta \\
0 & 1 & 0 \\
- \sin \eta & 0 & \cos \eta \end{array} \right ] . \] 
Thus, from the equation about for $\Delta v_z$, the
vertical heating can be estimated as a function of
$\eta$ and is
\begin{equation}
\Delta v_z =  - {{2GM_{pert}}\over{b^2 V_{sl}}} \, 
r \cos \phi_0 \,
\Biggl \lbrace {1\over 2} \alpha ^2 K_0(\alpha) 
\sin 2\eta \, + \, \alpha K_1(\alpha) \sin 2 \eta
\, \pm \, \alpha^2 K_1(\alpha) \sin \eta
\Biggr \rbrace .  
\end{equation}
It is straightforward to show that the dependence on
$\eta$ in this expression is such that the response
leads to a vertical warping of the disk, with an
efficiency depending on $\eta$.

\subsection{Parabolic orbits}

We now adopt the coplanar trajectory given in eqs. 
(\ref{polorb}) - (\ref{timeorb}) and apply the 
rotation matrix $\tilde{A}$ to the orbit.  Following
a similar procedure as for the straight-line path,
the velocity perturbations can be
written in the form:
\begin{equation}\label{dvxparabnoncop}
\Delta v_x  =  - {{2GM_{pert}}\over{b^2 V_{0}}} \, r \,
\Biggl \lbrace \left [ 2 \cos \phi_0 - 3 A_x \right ]
I_{20}(\sqrt{2} \alpha ) \, - \,
3 B_x I_{22}(\sqrt{2} \alpha ) \, - \,
3 C_x I_{2-2}(\sqrt{2} \alpha ) \Biggr \rbrace
\end{equation}
\begin{equation}\label{dvyparabnoncop}
\Delta v_y  =  - {{2GM_{pert}}\over{b^2 V_{0}}} \, r \,
\Biggl \lbrace \left [ 2 \sin \phi_0 - 3 A_y \right ]
I_{20}(\sqrt{2} \alpha ) \, - \,
3 B_y I_{22}(\sqrt{2} \alpha ) \, - \,
3 C_y I_{2-2}(\sqrt{2} \alpha ) \Biggr \rbrace
\end{equation}
\begin{equation}\label{dvzparabnoncop}
\Delta v_z  =  - {{2GM_{pert}}\over{b^2 V_{0}}} \, r \,
\Biggl \lbrace  - 3 A_z
I_{20}(\sqrt{2} \alpha ) \, - \,
3 B_z I_{22}(\sqrt{2} \alpha ) \, - \,
3 C_z I_{2-2}(\sqrt{2} \alpha ) \Biggr \rbrace \, ,
\end{equation}
where the terms $A_x, A_y, A_z, B_x, B_y, B_z, C_x,
C_y$, and $C_z$ are given by:
\begin{equation}
A_x = (a_{11}^2 + a_{12}^{2}) \cos \phi_0 \, + \, (a_{11} a_{21} + a_{21} a_{22}) \sin \phi_0
\end{equation}
\begin{equation}
B_x = {1\over 2} \left [ (a_{11}^2 - a_{12}^2 \mp a_{11} a_{22}
\mp a_{21}a_{12} )\cos \phi_0 
\, + \, (a_{11} a_{21} - a_{12}a_{22} \pm 2a_{11} a_{12}) \sin \phi_0
\right ]
\end{equation}
\begin{equation}
C_x = {1\over 2} \left [ (a_{11}^2 - a_{12}^2 \pm a_{11} a_{22}
\pm a_{21}a_{12} )\cos \phi_0 
\, + \, (a_{11} a_{21} - a_{12}a_{22} \mp 2a_{11} a_{12}) \sin \phi_0
\right ]
\end{equation}
\begin{equation}
A_y = (a_{12}a_{22} + a_{11}a_{21}) \cos \phi_0 
\, + \, (a_{21}^2 + a_{22}^2)  \sin \phi_0
\end{equation}
\begin{equation}
B_y = {1\over 2} \left [ (a_{11}a_{21} - a_{12}a_{22} \mp 2 a_{21} a_{22} )\cos \phi_0 
\, + \, (a_{21}^2 - a_{22}^2 \pm a_{12} a_{21} \pm a_{22}a_{11}) \sin \phi_0
\right ]
\end{equation}
\begin{equation}
C_y = {1\over 2} \left [ (a_{11}a_{21} - a_{12}a_{22} \pm 2 a_{21} a_{22})
\cos \phi_0 
\, + \, (a_{21}^2 - a_{22}^2 \mp a_{12} a_{21}
\mp a_{22} a_{11}) \sin \phi_0
\right ]
\end{equation}
\begin{equation}
A_z = (a_{32}a_{12} + a_{11}a_{31}) \cos \phi_0 
\, + \, (a_{32}a_{22} + a_{31}a_{21})  \sin \phi_0
\end{equation}
\begin{equation}
B_z = {1\over 2} \left [ (a_{11}a_{31} - a_{32}a_{12} \mp a_{31} a_{22}
\mp a_{32}a_{21} )\cos \phi_0 
\, + \, (a_{31}a_{21} - a_{32}a_{22} \pm a_{31} a_{12} \pm a_{32}a_{11}) \sin \phi_0
\right ]
\end{equation}
\begin{equation}
C_z = {1\over 2} \left [ (a_{11}a_{31} - a_{32}a_{12} \pm  a_{31} a_{22}
\pm  a_{32} a_{21})
\cos \phi_0 
\, + \, (a_{31}a_{21} - a_{32}a_{22} \mp a_{31} a_{12}
\mp a_{32} a_{11}) \sin \phi_0
\right ] .
\end{equation}
As earlier, the factors $a_{ij}$ are the 
relevant components of the rotation matrix $\tilde{A}$,
and the upper signs in the $\mp$ and $\pm$ terms refer to
the case with $\Omega > 0$ while the lower signs are for the
case with $\Omega < 0$.  In particular, whether the
encounter is mainly prograde or retrograde is set by
the sign of $\beta$, as given above.

In the equations for $\Delta v_k$, the functions
$I_{22}$ and $I_{2-2}$ can be simplified to expressions
involving $I_{l0}$ using the recursion relations
eqs.  (\ref{Rec1}) and (\ref{Rec2}).

\section{Numerical Experiments}

To test the reliability of the quasi-resonant approximation, we carry
out simulations of encounters between a spinning system and an
external perturber passing either on a straight line path or a
parabolic orbit.  For simplicity, we adopt a restricted three-body
method for following the development of tidal tails during collisions.
The restricted three-body scheme is appropriate for this application
because the formation of tidal tails is essentially a kinematic process
\citep{DMH99}.

In our experiments, the victim is a point mass surrounded by a flat
annular disk of test particles.  Initially, the particles are on circular,
Keplerian orbits around the central point mass.  
We employ a system of units in which Newton's constant, $G$,
and the maximum radius and mass of the victim are all set equal to unity.  

For each victim, we
place test particles on 400 annuli surrounding the point mass.  The
annuli are linearly spaced in radius, between $r_{\rm min}=0.05$ and
$r_{\rm max}=1$ around the central mass point with a total number of
approximately 1.3$\times$10$^{5}$ particles.  The annuli have an angular
spacing such that the particles are equidistantly distributed along
each ring, with a spacing $\Delta l=0.01=r \Delta\phi$. This choice
ensures a fair sampling of the outermost rings, a better visual
comparison, and an accurate sampling of the energy distribution of the
stars in the victim.  The perturber is modeled as a point mass passing
either on a straight line or parabolic orbit.  In this way, the
problem of the encounter is restricted to a three-body problem since
each test particle (``star'') feels only the gravitational force from
the massive particles (the central point mass of the victim and the
perturber).  We evolve the models for many dynamical times and compute
the energy distribution of the particles long
after the passage of the perturber, at a time where the energy
configuration of the system reaches a steady state. These
estimates are compared to analytical estimates using the
approximations presented earlier.

The comparison to the quasi-resonant impulse gained from each star in
the victim is done as follows.  We mimic the effect of encounters by
setting up the victim as a point mass surrounded by a flat annular
disk of particles.  We then assign to each test particle a circular,
Keplerian velocity plus a velocity increment estimated from the
analytic approximation for a given perturber mass and impact
parameter, and allow the disk to evolve.  The total energy of individual test 
particles is thus always conserved. While
our formalism can describe encounters of any mass ratio, we focus here
on equal mass situations with $M_{pert}=M_{victim}=1$, and an impact
parameter $b$=2, which is twice as large as our disk of test
particles.  For the straight line case, the choice of the relative
velocity $V_{sl}$ is arbitrary, so first we test our approximations
for a fast encounter (i.e. a weak perturbation), which is the case where our
formalism is expected to be most accurate.  In addition, we test
the validity of the approximations for slow encounters, 
which characterize stronger perturbations that are less impulsive.
 
In what follows, we present numerical tests of coplanar, straight line
encounters with fast and slow relative velocities, as well as
parabolic encounters at linear and second order approximations.  
The non-coplanar case has been also tested to
ensure that our analytic formulation is reasonable, but for brevity we
present results only for co-planar encounters. More general tests of
the validity of our analytic treatment will be considered in due
course in applications to observed systems.

\subsection{Results}

\begin{figure}[ht]
\epsscale{1.0}
\plotone{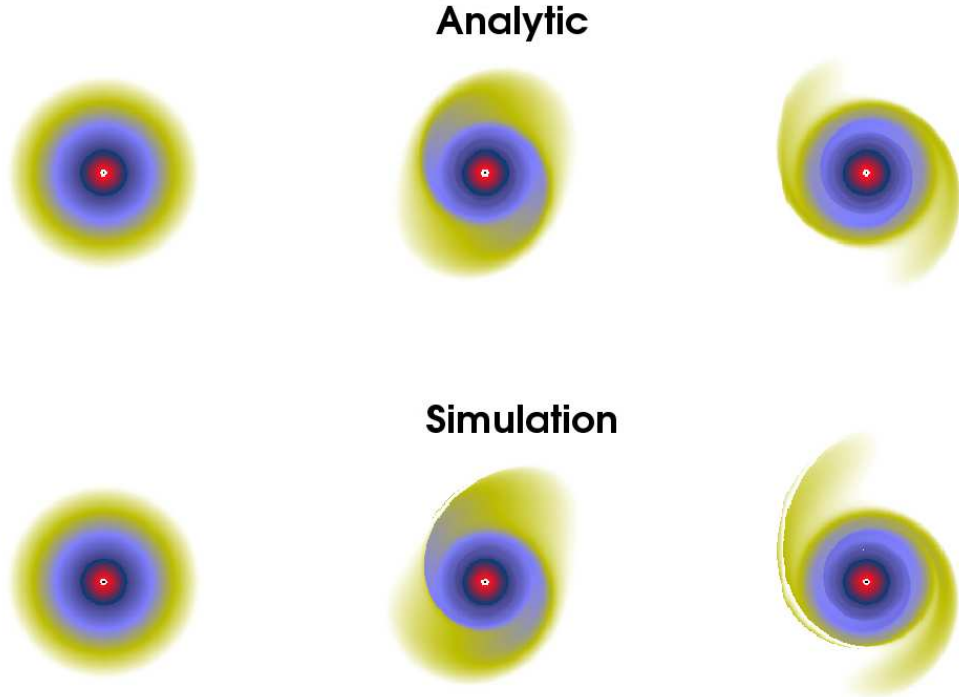}
\caption{Time evolution of the victim in the {\it x-y} 
plane under the tidal effect of a fast perturber passing on a straight-line path
on a {\it prograde} coplanar orbit.
The upper left panel illustrates the initial set up where test particles in the annuli are {\it kicked} by
velocity increments according to the quasi-resonant
approximation to linear order, as described in the text. Time evolves from left to right.
Bottom panels show the evolution in time of the victim (from the initial set up on the left to the 
final appearance of tails to the right) when an external perturber 
is passing on a straight line path, in a restricted three-body simulation.
The external perturber passes to the right side of the victim.
Colors are assigned to rings according to their initial distance from the central point mass.
The left panel shows the initial conditions, the middle panel is the system at time 3.5 and the right panel 
is the system at time 9.75.
\label{Tail_copl_pro}}
\end{figure}

\begin{figure}[ht]
\epsscale{1.0}
\plotone{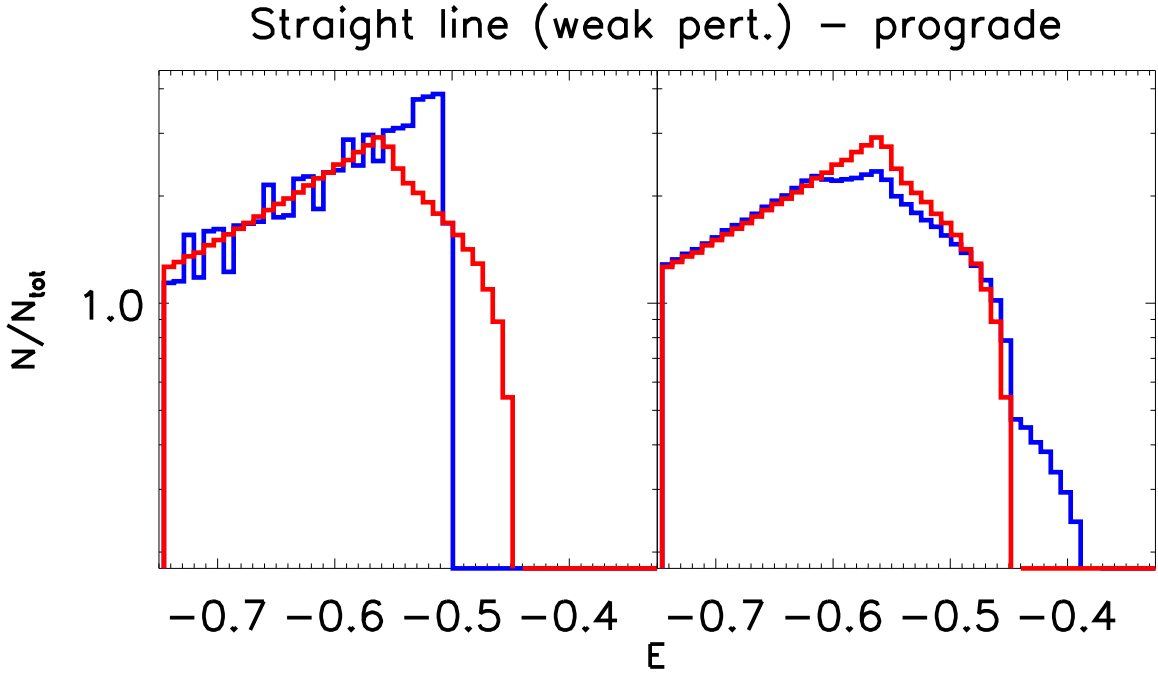}
\caption{Time evolution of the energy distribution of the stars of the victim 
long after the passage of the perturber on a fast straight-line path. The energy
of the victim particles in a simulation when an external perturber is involved (blue solid lines) is compared
to the energy distribution predicted by the quasi-resonant approximation to linear order (red solid lines).
In particular the red solid line is a setup of stars on Keplerian orbits with an
(initial) velocity distribution assigned according to the increment to the 
particle velocity given from our approximations. Note that the two red curves on the 
left and on the right panel are identical since the energy
of each test particle is strictly conserved along its orbit in this case.
The left panel shows the initial conditions and the right panel is the system at time 9.75. The time of closest approach 
in the simulation is 2.
\label{Epro}}
\end{figure}

\begin{figure}[ht]
\epsscale{1.0}
\plotone{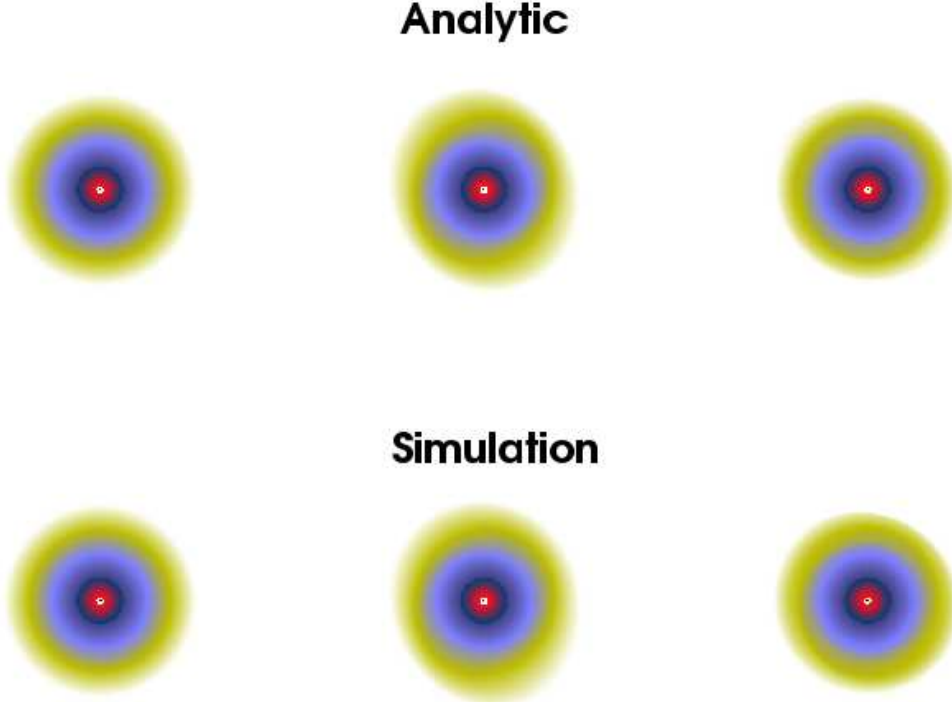}
\caption{Time evolution of the victim in the {\it x-y} plane 
under the tidal effect of a perturber passing on a straight-line path
on a {\it retrograde} coplanar orbit. Top and bottom panels are as in Figure \ref{Tail_copl_pro}.
The left panel shows the initial conditions, the middle panel is the system at time 3.5 and the right 
panel is the system at time 9.75.
\label{Tail_copl_retro}}
\end{figure}

\begin{figure}[ht]
\epsscale{1.0}
\plotone{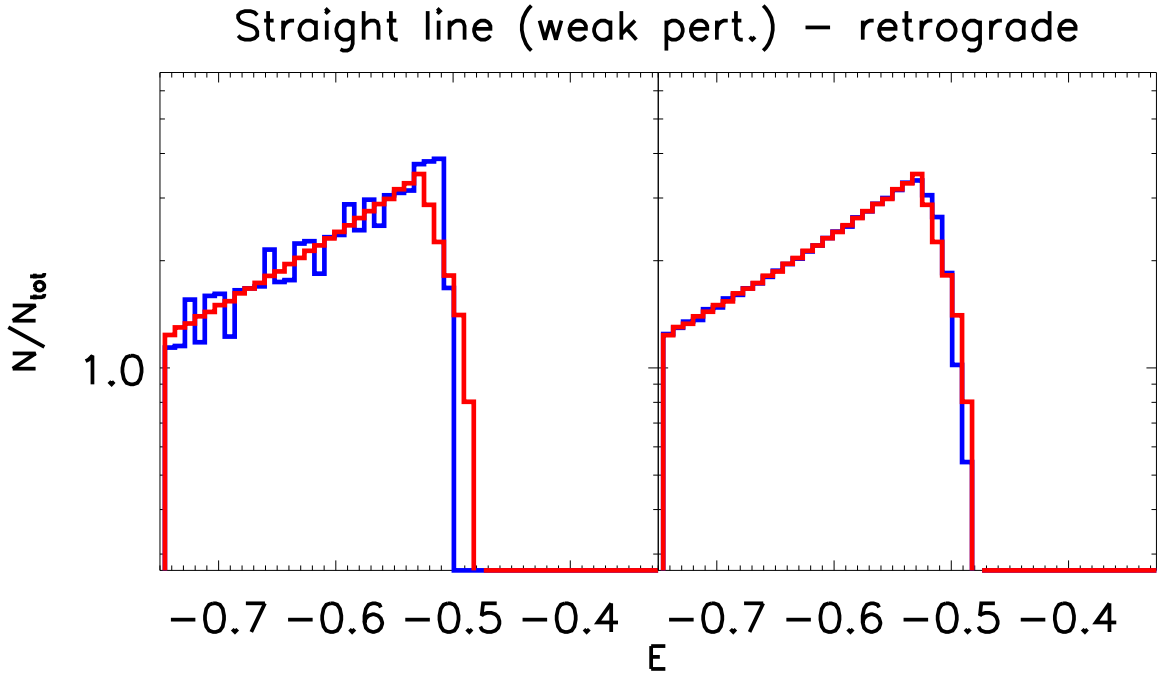}
\caption{Time evolution of the energy distribution of the stars in the victim 
long after the passage  of the perturber on a fast straight-line path on a {\it retrograde}  
orbit. The energy distribution of the particles perturbed by the linear order tidal approximation
is shown by the red solid lines, whereas the simulation with an external perturber is 
represented by the blue solid lines. The left panel shows the initial conditions and right panel is the system 
at time 9.75. The time of closest approach is 2.
\label{Eretro}}
\end{figure}

Figure \ref{Tail_copl_pro} gives a visual comparison of the evolution
of a victim under the tidal influence of the fast passage of a
perturber on a straight line with relative velocity $V_{sl}$=7.5.  The
top panels illustrate the system with test particles {\it kicked} by
velocity increments according to the quasi-resonant approximation
described in the previous sections.  The bottom panels display the
evolution of the victim when an external perturber passes close to the
disk using a restricted three-body treatment.  Figures
\ref{Tail_copl_pro}-\ref{Epro} illustrate a coplanar prograde encounter, while
the corresponding retrograde coplanar case is displayed in Figures
\ref{Tail_copl_retro}-\ref{Eretro}.  Colors are assigned to rings according to
their initial distance from the central point mass.

First, as expected, tails are produced for prograde encounters (Figure
\ref{Tail_copl_pro}), but are suppressed in retrograde cases (Figure
\ref{Tail_copl_retro}).  The visual agreement in the extents and
shapes of the tails between the analytic (upper panels) and the
numerical (lower panels) approaches is good.  Further insight is
provided by comparing the distribution of specific energies of the
disk particles as shown in Figures \ref{Epro} and Figure \ref{Eretro} for the
prograde and retrograde cases, respectively.
The different panels show the time evolution of the specific energies
for the restricted three-body simulation (blue) and the analytic
calculation (red), which is static.

\begin{figure}[ht]
\epsscale{1.0}
\plotone{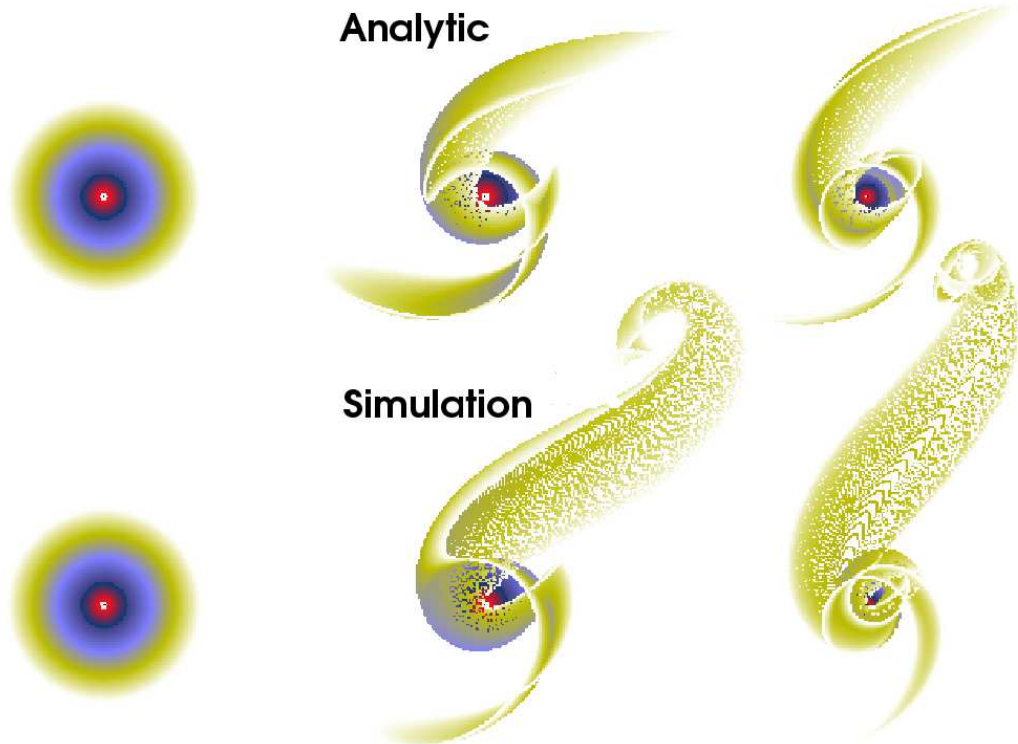}
\caption{Time evolution of the victim under the tidal effect of a slow perturber passing on a straight line path.
The orbit is {\it prograde}.
Top and bottom panels are as in Figure \ref{Tail_copl_pro}.
The left panel shows the initial conditions, the middle panel is the system at time 10.75 and  the right panel is the system 
at time 30.
\label{Tail_copl_pro_corr_sl}}
\end{figure}

Fast encounters correspond to relatively modest tidal perturbations,
leading to a weak resonant response.  The amount of energy transferred
during the encounter is small and the resulting tails are less
spatially extended than ones produced for stronger tidal
perturbations (see Figure \ref{Tail_copl_pro_corr_sl}, discussed below).    
As expected, prograde and retrograde encounters lead to very
different energy distributions, since prograde encounters transfer
much more energy to the victim than retrograde ones, producing much more
massive tails in the prograde case, as demonstrated in Figures 
\ref{Tail_copl_pro} and \ref{Tail_copl_retro}.  Indeed, according to our analytic results the maximum
resonant response is obtained for rings placed at distances
corresponding to the parameter $\alpha \sim 1$ (Figure \ref{alpha}).  In
this particular example, the outermost rings which contribute
significant material to the tails are only weakly resonant because
their spin frequencies are such that they have a value of $\alpha \sim 0.2 -
0.3$.  Note that our lowest-order approximations yield symmetric tails
as shown in the top panels of Figure \ref{Tail_copl_pro}, whereas the
numerical simulation yields tails that are slightly asymmetric (bottom
panels in Figure \ref{Tail_copl_pro}).  This effect increases for
stronger perturbations, and is connected to higher order corrections
as described in the sections above.

For retrograde encounters, the orbital and spin angular momenta are
misaligned and, as shown in Figure \ref{Eretro} for the evolution of
the energy distribution, little energy is transferred during the
encounter.  A resonance does not develop and tail formation is suppressed,
as shown in Figure \ref{Tail_copl_retro}.

The quasi-resonant approximation at lowest order, 
eq. (\ref{impulse}), becomes less accurate for slow encounters, where
the tidal forces are less impulsive and where the effective duration
of the encounter is of the order of the dynamical time of the
particles (``stars'') in the outermost rings.  Indeed, this is the
case where the restricted three-body simulation shows that some mass
from the victim is captured by the perturber, introducing a large
asymmetry and greater changes in the energy configuration of the final
system.  The lack of symmetry in tail-making owes to 
non-linear effects.  We can account for some of these by including the
next order correction to the response, as given by eqs.
(\ref{impulse_corr_sl_x}-\ref{impulse_corr_sl_y}) and
eqs. (\ref{impulse_corr_par_x}-\ref{impulse_corr_par_y}) for
the straight-line and parabolic cases, respectively.  We emphasize,
however, that even when we include these corrections we assume that
stars in the victim move along unperturbed orbits throughout the
course of the encounter, thereby treating the orbital variations as
small and implicitly ignoring the non-linear response.
 
\begin{figure}[ht]
\epsscale{1.0}
\plotone{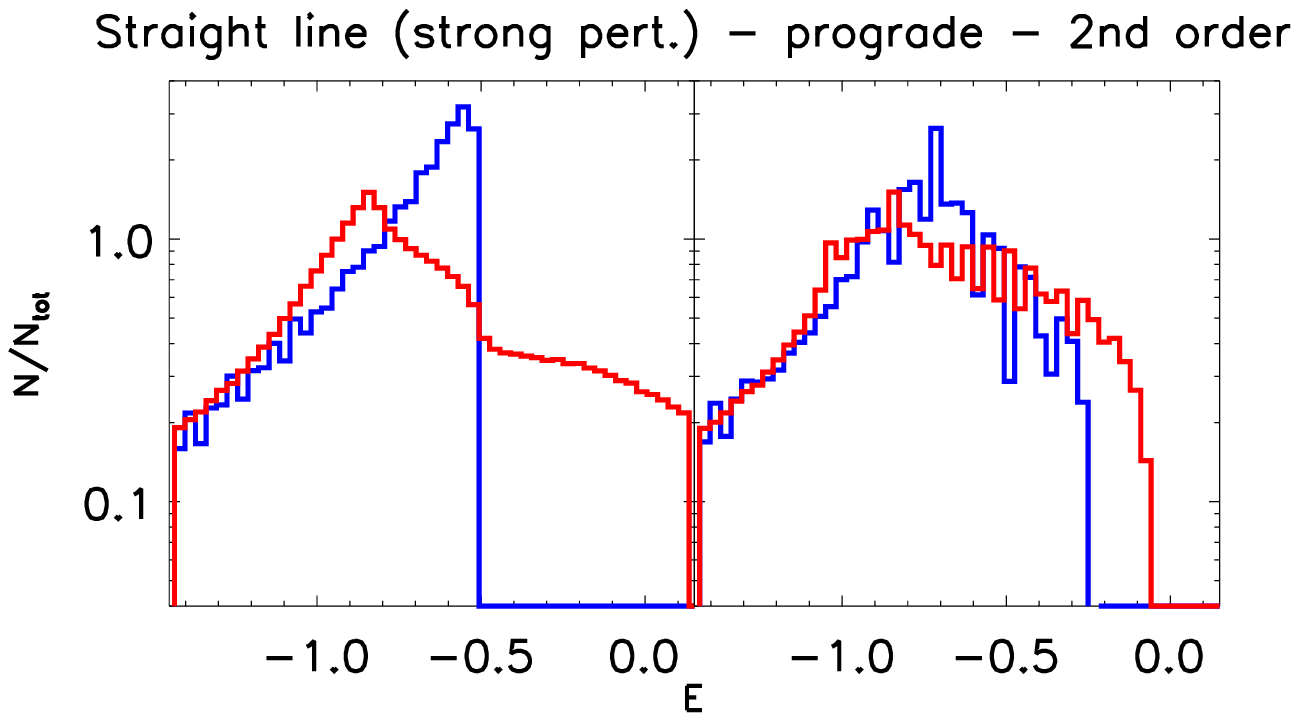}
\caption{Time evolution of the energy distribution of stars in the victim 
long after the passage of a perturber on a slow straight line path on a {\it prograde} orbit. 
The energy distribution of the particles perturbed by the second order  tidal approximation
is shown by the red solid lines, whereas the 
restricted three-body simulation of an encounter with an external perturber is 
represented by the blue solid lines. Since the distributions only involve the particles in the trailing
tails (y$<$0 in the {\it x-y} plane) the two red solid curves on the left and on the right panel are not identical.
The left panel shows the initial conditions and the right panel is the system 
at time 30. The time of closest approach 
in the simulation is 10.  
\label{En_sl_slow}}
\end{figure}

Figure \ref{Tail_copl_pro_corr_sl} shows a prograde straight-line
encounter for the relative velocity $V_{sl}=1.5$, as predicted by our
analytic approximation when the second order correction is included 
in the evaluation of the velocity increments given to the disk
particles (top panels). This evolution is compared to a simulation
with the same disk perturbed by an external body (bottom panels).  As
noted above, the tail connecting the victim to the perturber is not 
reproduced by our analytic formalism in detail because we ignore
the non-linear response of the victim during the encounter.
Nevertheless, the trailing tail, displayed in the region of the plane
with $y<0$, is similar in shape and orientation to the one generated
in the restricted three-body simulation.  Indeed our analytic
expression predicts the correct shape of the tail, although it is
slightly more extended compared to the one in the actual simulation.
The energy distribution of the particles contained in the trailing
tail in both models is shown in Figure \ref{En_sl_slow}, where the 
energy is shown in a logarithmic scale. 
Allowing for
the fact that our analytic method cannot entirely account for
non-linear effects, we find the comparison encouraging. 
Indeed, as the energy scale in Figure \ref{En_sl_slow}
is logarithmic, the regions showing obvious disagreement
contain few stars.

\begin{figure}[ht]
\epsscale{1.0}
\plotone{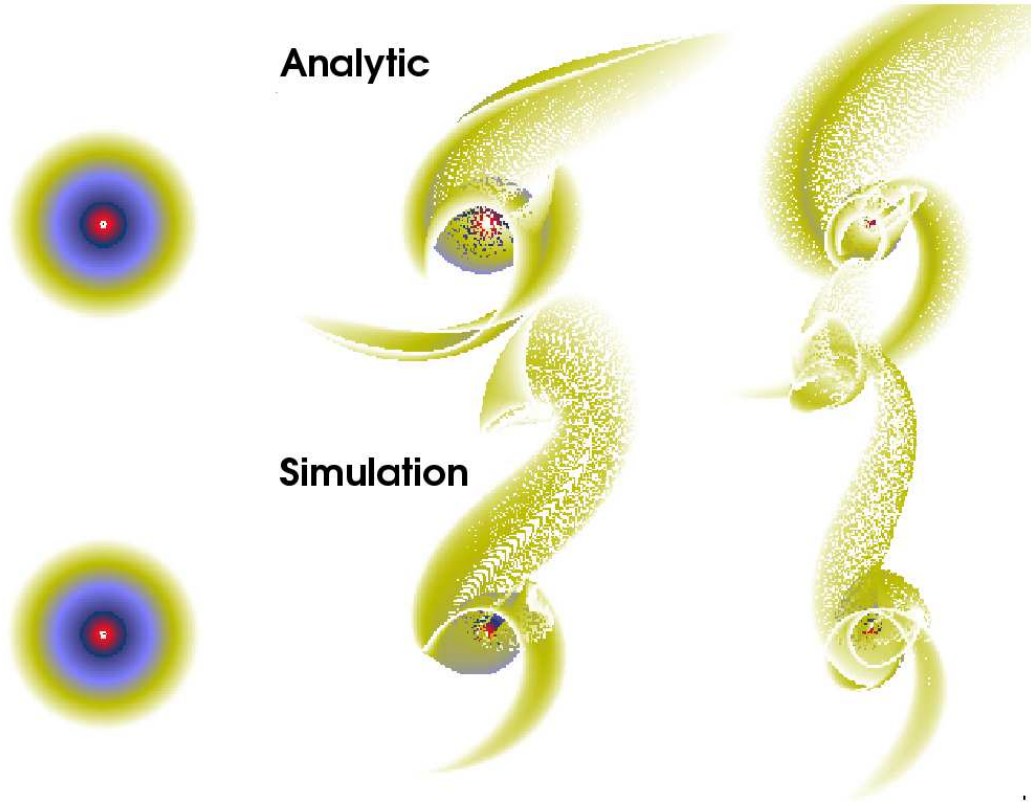}
\caption{Time evolution of the victim under the tidal effect of a perturber passing on a prograde 
coplanar parabolic orbit.
Top and bottom panels are as in Figure \ref{Tail_copl_pro}.
\label{Tail_copl_pro_par}}
\end{figure}

\begin{figure}[ht]
\epsscale{1.0}
\plotone{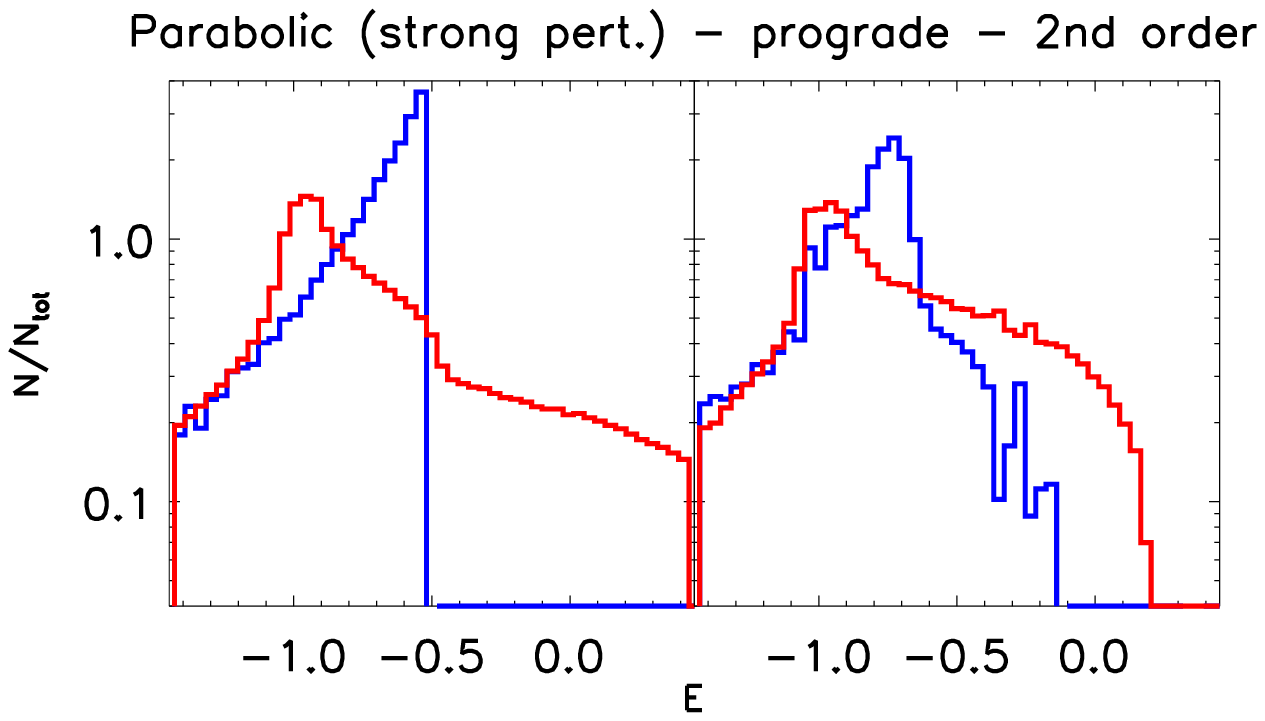}
\caption{Time evolution of the energy distribution of stars in the victim 
long after the passage of a perturber on a parabolic trajectory on a {\it prograde} orbit. 
The energy distribution of the particles using the analytical approximation to second order 
is shown by the red solid lines, whereas the encounter with an external perturber is 
represented by the blue solid lines.  Since the distributions only involve the particles in the trailing
tails (y$<$0 in the {\it x-y} plane) the two red solid curves on the left and on the right panel are not identical.
\label{En_par}}
\end{figure}

Next, we show an encounter from a parabolic trajectory where the
relative velocity of victim and perturber at closest approach is
$V_{0}=1.4$ for the parameters adopted in the
numerical experiments.  Figure \ref{Tail_copl_pro_par} shows a prograde
parabolic encounter as predicted by our analytic approximation when
the second order correction is included in the evaluation of the
velocity increments (top panels). The evolution is compared to a
restricted three-body simulation with the same disk as perturbed by an
external body passing on a parabolic trajectory (bottom panels).  This
encounter, as well as the slow straight line case, leads to strong
resonant effects.  In fact, the parameter $\alpha$, for both
cases, is around 1.3 for the outermost ring.  Hence these rings
should have a near-maximal response. This is demonstrated in
Figure \ref{Tail_copl_pro_par}, which shows longer tails as compared to 
the fast, straight-line encounter.  The asymmetry is also reflected in the
energy distribution of the slow straight-line encounter (Figure
\ref{En_sl_slow}) and of the parabolic case (Figure \ref{En_par})
which appear to be bimodal.

\section{Conclusions}

\noindent

We have formulated a simple analytic theory describing the resonant
response of rotating objects to gravitational tidal perturbations. We
estimated the velocity perturbations on a spinning disk resulting from
an encounter with a system on either a straight-line or a parabolic
orbit.  Our main conclusions are the following:

\begin{itemize} 
\item The resonant phenomenon described here is an incomplete or
 ``quasi-'' resonant process because the perturbation is applied
 at a given frequency for only a finite period of time.  
 Consequently, the resonance is broad and 
 is maximal when
 the spin frequency of the rotating victim matches the 
 peak orbital
 frequency of the encounter.  The resonance is broad because the
 perturbation at each frequency is not applied over many cycles,
 as for a strictly periodic forcing, so perturbations at 
 slightly off-resonant frequencies do not cancel out completely.
 However, because the resonance is broad, 
 at each point along the orbit the
 orbital frequency can be in resonance with parts of the victim disk
 located at different radii and can therefore act on
 various locations in the disk at different times.  This feature
 is interesting for galaxy encounters because it
 means that the victim can loose baryonic mass from different
 points in the disk by resonant stripping at various times,
 making the ``quasi-resonant'' response more damaging than
 if only certain regions in the disk were being affected.

\item In the quasi-resonant regime, the velocity increments gained by
the stars in the victim are greater during a parabolic encounter than
in the case of a straight-line interaction.  For non-rotating systems,
the velocity perturbations are described by the usual impulse
approximation.  In prograde cases (both straight or parabolic) the
velocity perturbations owing to quasi-resonant phenomenon are at
least factor $\sim 2$ higher than in the impulse approximation
while the response is suppressed for retrograde encounters.

\item When compared to numerical simulations we find that the
quasi-resonant approximation provides a simple physical model for
describing the nature of the tails in interacting disk galaxies
and gives well-defined and reasonable values for the changes
in velocities as a result of tidal encounters.  In particular, the
quasi-resonant approximation gives accurate results for high speed
encounters, where the effective duration of the encounter is less
than or of order the dynamical time of the particles (``stars'') of the spinning
system.  The main reason is that high velocity
encounters generate small perturbations of otherwise steady-state
systems and keep them in the linear regime, where the analytic
approximation is valid.   The quasi-resonant
approximation is less accurate for slower encounters,
where the tidal forces act over a longer period of time.
This is particularly true for the case of a
perturber passing on a parabolic trajectory.  Indeed, in these
events 
the numerical simulations show that some
mass of the victim galaxy is captured by the perturber.  
However, a better match can be obtained
by including higher order corrections for the velocity
perturbations in the analytic formalism.

\item The lengths of the tails depends on the amount of energy
transferred to the victim during an encounter.  Prograde encounters
with more massive perturbers pump larger amounts of energy 
into the victims, yielding more
extended tails but also introducing asymmetries with the leading tails
being more elongated than the trailing ones.  Retrograde encounters
transfer much less energy into the victim,
suppressing the development of tails.

\end{itemize}
\noindent
Finally, we note that our findings for tail-making
during galaxy-galaxy encounters may apply to gas as well. Thus, our calculations may be used 
to understand the geometry of the gas streams and bridges of stars
associated with dwarf galaxies, as  
recently discovered in the Panda survey of M31 \citep{McConn09}. Other 
potential applications include 
warps and heating of galactic disks by tidal perturbations,
understanding the conditions for long tails to be produced in
galaxy collisions depending on the dark matter distribution in
the halos, and identifying situations where stars can be unbound
in galaxy interactions and be ejected into the intergalactic
medium or populate the field of groups and clusters.  
Moreover, our
theory may be relevant to other applications, such as to
studies of the stability of binary stars  
perturbed by a third body or to
investigations
of protoplanetary disks perturbed by close passages of stars.

\acknowledgments 
\noindent
We thank Alar Toomre and Avi Loeb for valuable advice.  
ED acknowledges support from the Keck Foundation. 
CAFG is supported by a fellowship from
the Miller Institute for Basic Research in Science, and received further support from
the Harvard Merit Fellowship and FQRNT during the course of this work.

\appendix

\section{Generalized Airy functions}

Our analysis of parabolic tidal encounters results in expressions that
involve the generalized Airy functions of \citet{PT77},
defined by eq. (\ref{Ilmpt77}).
\noindent
The first one, $I_{00}$, is a modified Bessel function or 
(conventional) Airy function,
\begin{equation}
I_{00}(\sqrt{2}\alpha)=3^{-1/2}K_{1/3}\Big(2^{3/2}\frac{\sqrt{2}\alpha}{3}\Big)=\pi(2\alpha)^{-1/3}\rm{Ai}[(2\alpha)^{2/3}] \, .
\label{Ioo}
\end{equation}
\noindent
\citet{PT77} provide the following rational function
approximations to
$I_{10},~I_{20},$ and $I_{30}$, making it possible
to straightforwardly estimate the velocity
perturbations in a parabolic tidal encounter:

\begin{eqnarray}
I_{10}(y)
=\frac{1.5288+0.79192 \sqrt{y}-0.86606 y+0.14593 y\sqrt{y}}
{1.+1.6449\sqrt{y}-1.2345 y+0.19392 y\sqrt{y}}
e^{-\frac{2^{3/2}}{3}y},    \ \ \ \  \ \ \ \  \ \ \ \ \rm{for}  \ \ y \ \ \leq 4  \ \ \ \  \ \ \ \ \ 
\end{eqnarray}

\begin{eqnarray}
I_{10}(y)=\frac{1.4119+18.158 \sqrt{y}+22.152 y}{1.+12.249 \sqrt{y}+28.593 y}  
                               e^{-\frac{2^{3/2}}{3}y},   \ \ \ \   \ \ \ \  \ \ \ \  \ \ \ \ \rm{for}  \ \ y \ \  \geq 4 \ \ \ \  \ \ \ \ \ \ 
\end{eqnarray}

\begin{eqnarray}
I_{20}(y) 
=\frac{0.78374+1.5039 \sqrt{y}+1.0073 y+0.71115 y \sqrt{y}}{1.+1.9128 \sqrt{y}+1.0384 y+1.2883 y \sqrt{y}} 
\Big(1+\frac{2^{3/2}}{3}y \Big)^{1/2} e^{-\frac{2^{3/2}}{3}y}, \ \ \ \  \ \ \ \ \ \
\end{eqnarray}

\begin{eqnarray}
I_{30}(y)
=\frac{0.58894+0.32381 \sqrt{y}+0.45605 y+0.15220 y \sqrt{y}}{1.+0.54766 \sqrt{y}+0.76130 y+0.53016 y \sqrt{y}} 
                       \Big(1+\frac{2^{3/2}}{3}y \Big) e^{-\frac{2^{3/2}}{3}y}. \ \ \ \  \ \ \ \ \ \
\end{eqnarray}
These expressions are 
accurate to around $\leq 0.1$\%, except for very small and very large values of 
$\alpha$, where the error rises to a few percent.

To evaluate the limit $\alpha \to \infty$ of the parabolic encounter case studied in this work, it is necessary to know the asymptotic behavior of the generalized Airy functions as $y \to \infty$. 
In what follows, we give these asymptotic expansions and outline the key steps of their derivation.
While the $l=0$ case corresponds to the usual Airy function, the $l=1,~2,~3$ cases have not been extensively studied before, and the numerical fits provided by \citet{PT77} are not sufficiently accurate to capture exact cancellations in the results. 
\cite{O94} previously calculated these asymptotic expansions to leading order using similar techniques, but some of the limits we take involve exact cancellations at this order, so that it is necessary to compute higher-order terms. 

The starting point for the asymptotic expansions is the integral expression (eq. \ref{Ilmpt77})
\begin{equation}
\label{I l0 integral}
I_{l0}(y) = 
\int_{0}^{\infty} 
(1 + \xi^{2})^{-l} 
\cos{[\sqrt{2}y(\xi + \xi^3/3)]}
d\xi.
\end{equation}
The main technical difficulty in determining the asymptotic behavior of $I_{l0}(y)$ arises from the fact that the cosine term oscillates extremely rapidly as $y \to \infty$, and that the value of the integral depends on exactly how the positive and negative parts cancel each other. 
It is however possible to circumvent this difficulty by expressing equation (\ref{I l0 integral}) in terms of a contour integral in the complex plane, and by using the method of steepest descent \citep[e.g.,][]{1978amms.book.....B}.

\begin{figure}
\epsscale{0.5}
\plotone{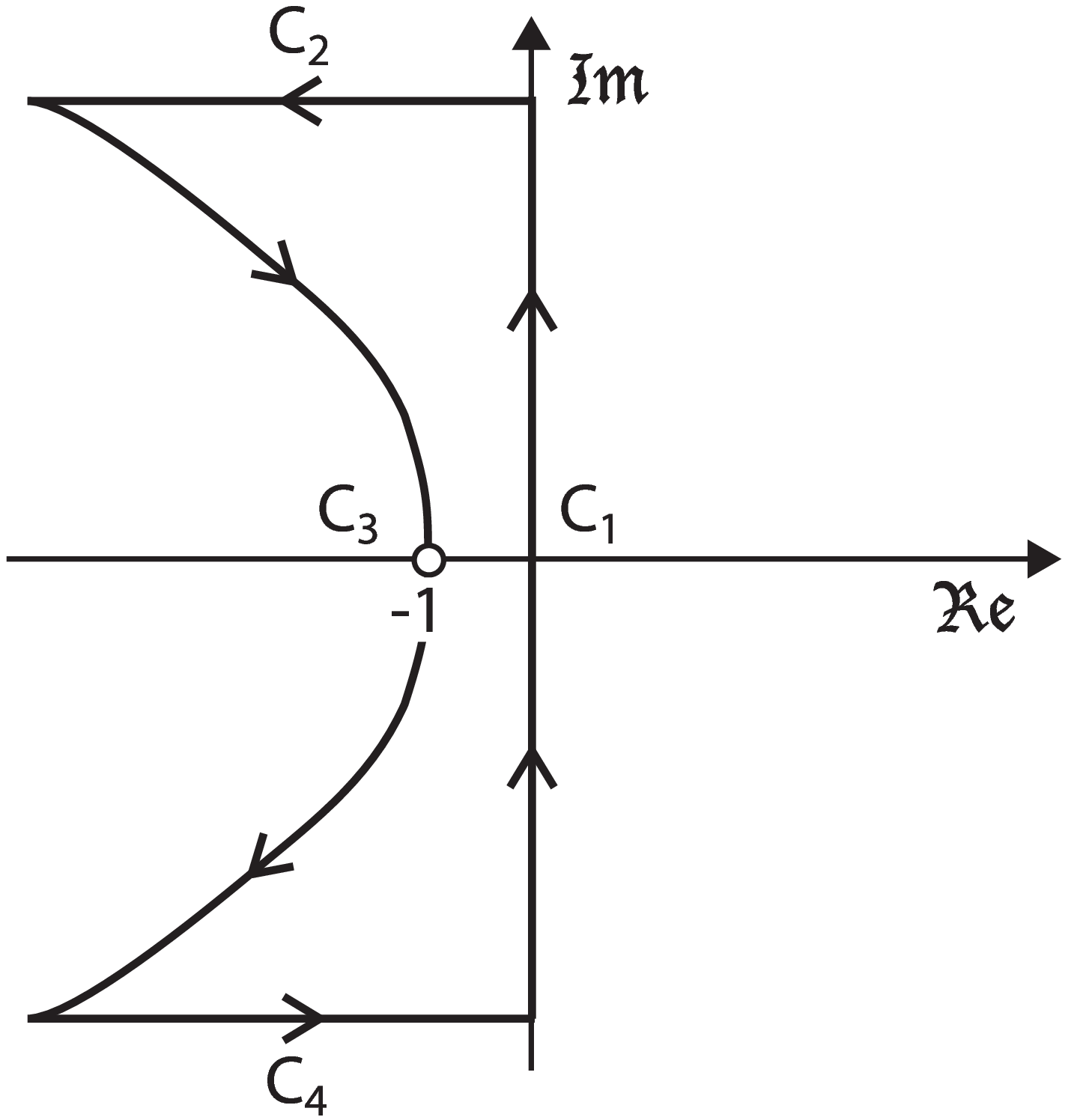}
\caption{Integration contour used in the asymptotic expansions of the generalized Airy functions $I_{l0}(y)$ for $y \to \infty$. 
$C_{3}$ corresponds to the path of steepest descent through the saddle point $s=-1$ of the integral in equation (\ref{I l 0 after change of var}). 
The method of steepest descent stipulates that most of the contribution to the integral along this path comes from the neighborhood of this saddle point as $y \to \infty$. 
Using the residue theorem, the desired integrals along the imaginary axis ($C_{1}$) can be evaluated asymptotically.}
\label{contour}
\end{figure}

Specifically, we note that
\begin{equation}
I_{l0}(y) = 
\frac{1}{2}
{\rm Re} \left\{
\int_{-\infty}^{\infty} 
(1 + \xi^{2})^{-l} 
\exp{[i\sqrt{2}y(\xi + \xi^3/3)]}
d\xi
\right\}
\end{equation}
and consider the integration contour shown in Figure \ref{contour}.
Making the change of variable $s\equiv i \xi$, the integral within the curly brackets is seen to be equal to an equivalent integral along the imaginary axis:
\begin{equation}
\label{I l 0 after change of var}
-i \int_{-i \infty}^{i \infty} (1-s^{2})^{-l} \exp{[\sqrt{2}y(s - s^{3}/3)]} ds.
\end{equation}
Writing the complex integrand as $f(s)$, the residue theorem implies that $\oint_{C} f(s) ds = 2 \pi i \sum_{k=1}^{n} {\rm Res}(f,~a_{k})$, where the $a_{k}$ are the poles within $C=C_{1}+C_{2}+C_{3}+C_{4}$. 
Since the integrands along $C_{2}$ and $C_{4}$ are exponentially suppressed in modulus, $\int_{C_{2},C_{4}}f(s)ds\to 0$ as the contour is expanded to infinity, yielding
\begin{equation}
\label{residue theorem applied}
\int_{C_{1}} f(s) ds = -\int_{C_{3}} f(s) ds + 2 \pi i \sum_{k=1}^{n} {\rm Res}(f,~a_{k}).
\end{equation}
As the left hand side is simply related to equation (\ref{I l 0 after change of var}) by the multiplicative factor $-i$, the problem is essentially reduced to evaluating $\int_{C_{3}}f(s)ds$.

The key is to choose the path $C_{3}$ so that the integral is tractable in the limit $y \to \infty$. 
According to the method of steepest descent, the integral will in this limit receive most of its contribution in the neighborhood of a saddle point, if the integration path is chosen to follow the direction of steepest descent of the exponential argument. 
Setting $d[s - s^{3}/3]/ds=0$, the relevant saddle point is found to be $s=-1$, and we may parametrize the path of steepest descent by 
\begin{equation}
\label{steepest descent path}
s - s^{3}/3 = -2/3 - t^{2},~t \in (-\infty, \infty), 
\end{equation}
chosen so that $s=-1$ at $t=0$.
Then,
\begin{equation}
\label{C3 t integral}
\int_{C_{3}} f(s) ds = \int_{-\infty}^{\infty} f(s(t)) \frac{ds}{dt} dt
= \exp{\left(-\frac{2^{3/2}}{3}y\right)} \int_{-\infty}^{\infty} \exp{\left(-\sqrt{2} y t^{2}\right)} \frac{1}{[1 - s(t)^2]^{l}} \frac{ds}{dt} dt,
\end{equation}
provided that the parametrization $s(t)$ is oriented in the same sense as $C_{3}$. 
Once an explicit expression for $s(t)$ is obtained by solving the cubic equation (\ref{steepest descent path}), we expand the product $[1 - s(t)^2]^{-l} (ds/dt)$ in a Taylor series around $t=0$. 
For instance, the $l=0$ case yields $ds/dt = i - (5i/24)i t^{2} + (385i/3456) t^{4} + ...~+$ purely real terms, where the purely real terms do not enter in the final result.
The integral on the right hand side of equation (\ref{C3 t integral}) becomes a term-by-term sum of simple Gaussian integrals of the form $b_{n} \int_{-\infty}^{\infty} t^{n} \exp{\left(-\sqrt{2} y t^{2}\right)} dt$, where $n=0,~1,~2,~3,~...$ and the $b_{n}$ coefficients are determined by the Taylor expansion. 
These integrals are easily evaluated and we thus obtain an analytic expansion for $\int_{C_{3}}f(s)ds$ valid in the limit $y \to \infty$.

This procedure can be carried out as described for $l=0$. 
In this case, there is no residue and we find
\begin{equation}
\label{I 0l infinity}
I_{00}(y) \sim 
\frac{\sqrt{\pi}}{2^{5/4}}
\exp{\left(-\frac{2^{3/2} }{3} y\right)}
\left[
\frac{1}{y^{1/2}} 
-
\frac{5}{48 \sqrt{2}}
\frac{1}{y^{3/2}}
+ 
\frac{385}{9216} 
\frac{1}{y^{5/2}}
+ 
O\left(\frac{1}{y^{7/2}}\right)
\right],
\end{equation}
in agreement with the known result for the ordinary Airy function. 
For $l>0$, we however encounter the difficulty that the saddle point $s=-1$ is also a pole of order $l$.

For the simple pole of the case $l=1$, the basic procedure remains valid but we must account for the ${\rm Res}(f,-1)$ contribution in equation (\ref{residue theorem applied}). 
Since the pole is located directly on the contour (rather than contained within it) it contributes only half of its value, namely $\pi i \times {\rm Res}(f,-1)$ rather than $2 \pi i \times {\rm Res}(f,-1)$, as can be shown by considering infinitesimally modified contours that avoid and enclose the pole. 
We find ${\rm Res}(f,-1) = (1/2) \exp{[-(2^{3/2}/3)y]}$ and 
\begin{equation}
\label{I 1l infinity}
I_{10}(y) \sim 
\frac{\pi}{4}
\exp{\left(-\frac{2^{3/2} }{3} y\right)}
\left[
1
+
\frac{2^{3/4}}{3\sqrt{\pi}}
\frac{1}{y^{1/2}}
-\frac{47}{216 \times 2^{3/4} \sqrt{\pi}}
\frac{1}{y^{3/2}}
+
O\left(\frac{1}{y^{5/2}}\right)
\right].
\end{equation}\\
For the $l=2$ and $l=3$ cases, the procedure yields divergent integrals. 
These can however be avoided by going back to equation (\ref{I l 0 after change of var}) and using a combination of integration by parts and partial fraction expansions to rewrite it in terms of $I_{00}(y)$, $I_{10}(y)$, and other integrals that are regular at $s=-1$ and everywhere within $C$. 
The latter integrals can then be evaluated in the limit $y \to \infty$ using the method of steepest descent, in close analogy to the $l=0$ case.
We finally find:
\begin{equation}
\label{I 2l infinity}
I_{20}(y) \sim
\exp{\left(-\frac{2^{3/2} }{3} y\right)}
\left[
\frac{\pi}{8}
+
\frac{23 \sqrt{\pi}}{192\times 2^{1/4}}
\frac{1}{y^{1/2}}
+
\frac{2869 \sqrt{\pi}}{55296\times 2^{3/4}}
\frac{1}{y^{3/2}}
+
O\left(\frac{1}{y^{5/2}}\right)
\right]
\end{equation}
and
\begin{equation}
\label{I 3l infinity}
I_{30}(y) \sim
\exp{\left(-\frac{2^{3/2} }{3} y\right)}
\left[
\frac{\pi}{8\times 2^{1/2}}
y
+
\frac{\sqrt{\pi}}{3\times 2^{3/4}}
y^{1/2}
+
\frac{3 \pi}{32}
+
O\left(\frac{1}{y^{1/2}}\right)
\right].
\end{equation}

\bibliographystyle{apj}

\end{document}